\documentclass{aastex63}

\usepackage{amsmath}
\usepackage{xcolor}
\journalinfo{Release Number: LLNL-JRNL-2011107}

\begin{document}

\title{$r$-Process Nucleosynthesis from Hyperaccreting Neutron Stars in Common Envelopes}

\author{Peter Anninos}
\affil{Lawrence Livermore National Laboratory, Livermore, CA 94550, USA}

\author{Matthew E. Portman}
\affil{ Space Telescope Science Institute, 3700 San Martin Drive, Baltimore, MD 21218, USA}

\author{Scott R. Carmichael}
\affil{Department of Physics and Astronomy, University of Notre Dame, Notre Dame, IN 46556}

\author{Robert D. Hoffman}
\affil{Lawrence Livermore National Laboratory, Livermore, CA 94550, USA}

\author{Andre Sieverding}
\affil{Lawrence Livermore National Laboratory, Livermore, CA 94550, USA}

\begin{abstract}
We investigate nuclear reactions and feedback in hyperaccreting neutron star environments, considering accretion rates in the range 0.3 - $3\times10^4$ $M_\odot$ yr$^{-1}$, typical of short-period compact object binaries in common envelopes. Our models account for weak reactions, neutrino energy loss, nuclear energy release, pair production, degenerate equations of state, and general relativistic hydrodynamics. Depending on accretion rates, these systems can develop both proton and neutron-rich atmospheres with strong convective instabilities linking the neutrino sphere to the outgoing accretion shock inside the radiation trapping zone. Convection drives nucleons through multiple heating and cooling cycles,  with photodisintegration dominating during the heating phase and heavy element synthesis during the cooling phase, ejecting material with abundances that depend on the accretion rate and depth of the final decompression trajectory. The turbulent nature of convective currents is conducive to creating a wide range of nuclear products through a variety of effects, including NSE freezeout and $r$, $p$ and $\gamma$ processes. We also observe a novel multi-step process in reheated trajectories, consisting of proton-capture and photo-dissociation reactions operating on $r$-process seeds, producing overall neutron-deficient isotopes. A significant amount of infalling gas experiences high entropy and short (millisecond) freezeout timescales capable of making $r$-process elements with high over-abundances through a disequilibrium effect between neutrons and 
$\alpha$-particles that does not require an excess of neutrons.
\end{abstract}


\section{Introduction}
\label{sec:intro}

A common envelope (CE) environment is known to form around compact-object binaries when accreting material fills the Roche Lobe and both objects are consumed by the overflow \citep{Paczynski76}. Orbital energy is lost during this process through dynamical and tidal friction, but also through various transport mechanisms, including gravitational wave, neutrino, and photon emissions, driving the two objects (and accreting material) spiraling towards one other. For the particular case of an expanding red giant - neutron star (NS) binary, the combination of high orbital velocities and gravitational pull of the NS draws material from the red giant at \replaced{hypercritical rates, resulting in the buildup of an atmosphere} {rates capable of creating atmospheres} dense and hot enough to support the creation of neutrinos and either proton or neutron rich elements. Eventually some of this material is ejected from the binary system, contributing heavy elements to the intergalactic medium \citep{Keegans19}. These dissipation mechanisms may also ultimately be responsible for the ejection of the entire common envelope, though exactly how or if such an outcome is possible remains in question \citep{Livio88,Taam00,Taam10,Ivanova13}.

During the early stages of a developing CE, an accretion shock forms on the surface of the neutron star, driven by infalling material, and moves outward to some relatively stable radius as the post-shock region settles into a pressure-equilibrated atmosphere \citep{Chevalier89,Houck91}. The process of equilibration can last for many dynamical times, and becomes impractical to simulate numerically in multi-dimensions if one additionally requires the spatial resolution to track nucleosynthesis and neutrino loss. Because our interest is on modeling the hot environment where heavy element production occurs, we emphasize the near-surface physics and do not simulate the entire binary system nor evolve long enough to completely settle the atmosphere. Instead we focus on resolving the dense shock-heated accretion atmosphere, the thin neutrino cooling layer that forms on the NS surface, and the early convective instability driving the shock and cyclic heavy element production.

We consider accretion rates 8 to 13 orders of magnitude greater than the Eddington rate, ranging from $\dot{M} = 10^8 \dot{M}_{Edd} \approx 0.3 ~M_\odot/\text{yr}$  to $\dot{M} = 10^{13} \dot{M}_{Edd} \approx 3\times10^4 ~M_\odot/\text{yr}$. At these rates, the optical depth is large enough to prevent photons from escaping (they are carried inward faster than they diffuse outward) so that much of the released accretion energy cannot be transported by photons. Energy is instead transported by neutrinos produced near the neutron star surface. As material flows toward the neutron star, inside the trapping radius, compression of the radiation pressure dominated gas increases the temperature above the pair production threshold. Pair annihilation produces neutrinos which immediately escape, carrying  energy without interacting with infalling material. (We assume the neutrino mean free path exceeds the problem domain\added{, as shown in Figure \ref{fig:opticaldepth},} so that neutrino transport can be ignored, an assumption we plan to revisit in future work). Neutrino losses trigger a cooling instability, allowing the neutron star to accrete well above the Bondi-Hoyle rate. The total accreted mass can be considerable, perhaps even enough for the neutron star to collapse to a black hole \citep{Fryer96}. Hence resolving the sub-kilometer neutrino cooling scale is critical to accurately model local behavior as well as any global implications to the CE system on the whole.

In multi-dimensions, the transient structure evolving behind the outward propagating accretion shock
becomes convectively unstable \citep{Woosley02,Bernal13,Fraija18}, as turbulent energy drives the shock well beyond its spherical steady state position. This creates conditions ripe for heavy element synthesis due to fluid parcels being rapidly heated then cooled cyclically through convective turnover. While neutron star mergers are one likely site for the production of heavy elements by the $r$ process \citep{Abbot:2017,Drout:2017,Kilpatrick:2017}, there is observational evidence that other sites are needed \citep{Cote.Eichler.ea:2019,Molero.Magrini.ea:2023}. But where such processes occur in nature remains speculative and uncertain \citep{Cote18}. Fall-back accretion in supernova explosions \citep{Qian96,Fryer06} is one of several such proposed sites that closely resembles the CE scenario considered here (see \cite{Freiburghaus99}, \cite{Lattimer74}, and \cite{Suzuki05} for others). Supernova fall-back relies on material accreting onto a (proto) neutron star to produce conditions appropriate for making heavy elements, but at significantly higher accretion rates (by roughly a few orders of magnitude) than expected from CEs \citep{Keegans19}.

Here we investigate both proton and neutron rich heavy element nucleosynthesis in convectively unstable atmospheres of hyper-accreting neutron stars in CEs. We present production yields from unbound matter, using population synthesis models to study the distribution of burn products across five decades of accretion rates.
Section \ref{sec:methods} begins with a brief discussion of our numerical methods, physical models, and gridding strategy tuned to achieve the desired high spatial resolution near the neutron star surface. Our 1D and 2D results follow in Sections \ref{sec:1Dresults} and \ref{sec:2Dresults} respecively, and we conclude with a brief summary in Section \ref{sec:summary}. Throughout we use standard values for the neutron star mass ($M_{NS}=1.4$ $M_\odot$) and radius ($R_{NS}=10$ km). Mass units are set by the NS mass, length units by its radius, and we adopt time units where the speed of light is unity.

\section{Methods and Models}
\label{sec:methods}

All calculations are performed with the {\sc Cosmos++} code \citep{Anninos05,Anninos17,Anninos20,Fragile14,Roth22}
which solves the general relativistic RMHD equations coupled with thermonuclear reactions, energy generation, and radiation transport.  {\sc Cosmos++} is a parallel, multi-dimensional, multi-physics, object-oriented code, supporting adaptive structured and unstructured meshes, finite volume and finite element discretization, and
for general relativistic (as well as Newtonian) astrophysical applications. For this work we utilize finite volume, high resolution Godunov shock capturing with third-order piecewise-parabolic interpolations for flux reconstructions, and third-order (low-storage) Euler time-stepping. Since all of the code attributes utilized in this report have been described (and vetted) in previous publications, we provide here only brief descriptions of relevant features.

\subsection{General Relativistic Hydrodynamics}
\label{subsec:hydro}

Although {\sc Cosmos++} is equipped to solve the general relativistic radiation-magnetohydrodynamic equations, we include neither radiation nor magnetic fields in these studies. Plasma  densities and temperatures at the hyper-Eddington rates considered here are sufficiently large that thermal radiation is trapped and advected with accreting flow, negating the need for radiation transport and allowing for the plasma to cool instead by decoupled neutrino emission. Additionally, neutron stars can have appreciable magnetic fields (in excess of $10^{12}$ G), but the resulting radiation pressure and kinetic infall energy of accreting material dominate over magnetic pressure, and effectively smother or bury field lines within the NS crust at the accretion rates considered in this report \citep{Chevalier89,Fryer96,Bernal13}.

Thermodynamics is treated with a Helmholtz equation of state, accounting for radiation, electron degeneracy, relativistic electron-positron plasma, and Coulomb corrections \citep{Timmes00}. Our implementation, based on the Torch code \citep{Timmes99}, is designed to work with arbitrary isotopic compositions and inline nuclear reaction networks.
It is utilized in tabular form with densities spanning $10^{-12} \le \rho \le 10^{15}$ g cm$^{-3}$, and temperatures $10^{3} \le T \le 10^{13}$ K. Interpolations between table entries are performed with biquintic Hermite polynomials that ensures thermodynamic consistency.

\subsection{Neutrino Cooling}
\label{subsec:neutrino}

\begin{figure}
\hspace{0.9in} \includegraphics[width=0.7\textwidth]{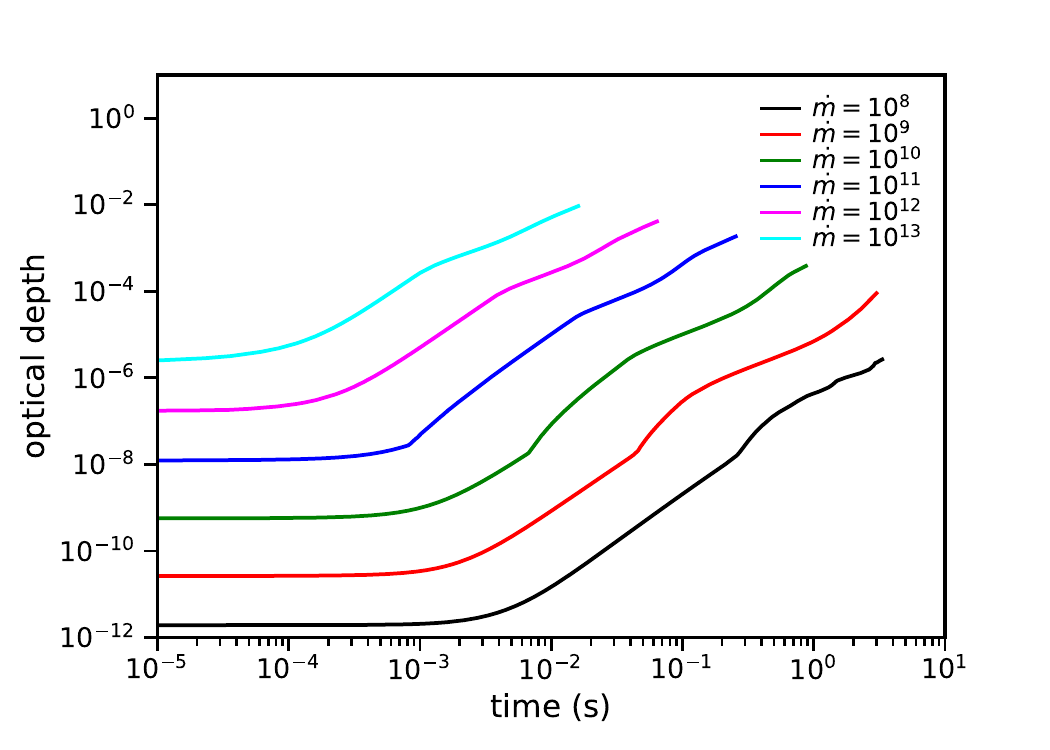}
\vspace{0.2in}\caption{
The neutrino optical depth attributed to elastic scattering is integrated radially across the computational domain and plotted as a function of time for all the calculations.
}
\label{fig:opticaldepth}
\vspace{0.2in}
\end{figure}

We account for five neutrino interactions, including pair annihilation, bremsstrahlung off free nucleons, photodisintegration, plasmon decay, and nucleon recombination \citep{Itoh89,Itoh96}. Cooling rates are pre-tabulated then interpolated as functions of temperature, density, and mean atomic weight and number. We assume that when neutrinos are born they immediately carry their energy beyond the boundaries of the problem, thus serving as pure energy-sinks in these models. As justification, we plot in Figure \ref{fig:opticaldepth} the neutrino optical depth across the entire radial domain as a function of time ($\tau = \int dr/\lambda_\nu$), taking elastic scattering for the the mean free path $\lambda_\nu$ \citep{Brown82}
\begin{equation}
\lambda_\nu = 10~\text{km} ~\rho_{12}^{-1} \left(\frac{\epsilon_\nu}{10~\text{MeV}}\right)^{-2}
              \left[\frac{\overline{N}^2}{6\overline{A}} ~x_h + x_n + \frac{5}{6} x_p \right]^{-1} ~,
\end{equation}
where $\rho_{12}$ is the density in units of $10^{12}$ g/cm$^3$, $\epsilon_\nu$ is the average neutrino energy (10 MeV), $\overline{N}$ is the average number of neutrons in heavy nuclei, $\overline{A}$ is the average number of nucleons, and $x_h$, $x_n$, $x_p$ are the mass fractions of heavy nuclei, neutrons, and protons respectively. Although the optical depth increases in time, it remains less than $10^{-2}$ by the end of each calculation even for the highest accretion rate models.

\subsection{Nuclear Networks}
\label{subsec:nucleo}

{\sc Cosmos++} supports a number of inline nuclear reaction networks, including 7-, 19- and 21-isotope $\alpha$-chain and heavy-ion reaction models \citep{Weaver78,Timmes99,Anninos18}, which, although lacking in reactant fidelity, captures nuclear energy production well enough across a range of stellar environments and burn conditions. All of the inline nucleosynthesis results are calculated with the same 19-isotope model used in our black-hole tidal disruption (TD) work \cite{Anninos18,Anninos19,Anninos22}, so we refer the reader to those references for further details regarding the network, implicit solvers, species advection, and relativistic energy coupling algorithms. Essentially, the 19-isotope network solves heavy-ion reactions using a standard $\alpha$-chain  composed of 13 nuclei ($^{4}$He, $^{12}$C, $^{16}$O, $^{20}$Ne, $^{24}$Mg, $^{28}$Si, $^{32}$S, $^{36}$Ar, $^{40}$Ca, $^{44}$Ti, $^{48}$Cr, $^{52}$Fe, and $^{56}$Ni), plus additional isotopes ($^{1}$H, $^{3}$He, $^{14}$N, $^{54}$Fe) to accommodate some types of hydrogen burning as well as photo-disintegration into neutrons and protons. \added{But as we point out in a later section, our results and conclusions are mostly insensitive to nuclear energy production.}

This work additionally includes weak reactions, accounting for neutrino fluxes, electron decay, electron capture, positron decay, and positron capture \citep{Langanke01}. Also, above a temperature threshold of 8 GK nuclear statistical equilibrium (NSE) is enforced via a Newton-Raphson scheme that converges on the electron fraction $Y_e$ after solving the reduced weak reactions network.

We utilize several options for more extensive and detailed post-processing of reactant histories: (1) expanded versions of Torch \citep{Timmes99}, (2) PRISM \citep{Mumpower17}, and (3) Winnet \citep{Reichert23} have all been adapted to the {\sc Cosmos++} framework and modified for parallel processing. Torch is used to validate inline yields from intermediate mass networks;  PRISM and Winnet to solve large heavy element networks appropriate for the $r$-process. Post-processing is performed using a 7150 isotope reaction network, with elements up to Rg$^{339}_{111}$, drip line to drip line. Rates are based on the Reaclib database \citep{Rauscher2000} including reactions with nucleons, photons, $\alpha$-particles, decays, neutrinos, and electron-positron captures. We also independently cross-check post-processed results using XNet \citep{Hix.Meyer:2006}.

\subsection{Neutron Star Crust}
\label{subsec:crust}

Rather than impose a rigid, perfectly reflecting boundary condition along the neutron star surface, we instead soften the boundary by building a hydrostatic outer crust up from the edge of the grid, extending the analytic equilibrium shock solution to provide estimates of surface densities and temperatures. Enforcing an isothermal profile throughout this
crust region allows us to iteratively solve the hydrostatic equation 
\begin{equation}
\frac{dP}{dr} = |g| ~\rho = \left(\frac{GM_{NS}}{R_{NS}^2}\right) \rho  ~,
\label{eqn:hydrostatic}
\end{equation}
converging (via Newton-Raphson or bisection) on the density and pressure solutions -
a procedure that works for an arbitrary equation of state \citep{Malone14}. Equation (\ref{eqn:hydrostatic}) is integrated outward from the boundary until the density (or pressure) falls below the free-fall solution. Typically the crust constructed in this manner ends up a small fraction of the computational grid, less than a percent of the NS radius, which we fill with iron-group nuclei. We additionally apply hydrostatic (not reflection) boundary conditions off the grid.

\subsection{Initial Data}
\label{subsec:initial}

If the density is uniform upstream beyond the gravitational influence of the neutron star,
the Bondi-Hoyle-Lyttleton (BHL) theory applies \citep{Hoyle39,Bondi44}, allowing us to estimate
the mass accretion rate as $\dot{M} \approx 4\pi r_B^2 \rho v$ with Bondi radius $r_B = GM_{NS}/v^2$, 
where $\rho$ and $v$ are the density and velocity of the upstream flow,
$M_{NS}$ is the mass of the accretor,
and we have neglected the sound speed in the usual writing of these expressions.

We begin each calculation with the fluid in spherically symmetric free-fall, using the BHL
formulation and mass conservation to write
\begin{equation}
u^r_{ff} = \sqrt{\frac{2GM_{NS}}{r}} ~,  \qquad \rho_{ff} = \frac{\dot{M}}{4\pi r^2_{sh}} \frac{1}{u^r_{ff}} ~,
\end{equation}
for the radial 4-velocity and mass density respectively, 
from which the coordinate velocity follows $v^r=u^r/u^0$, and the time component $u^0$
(the Lorentz factor in flat space) is derived from $u^\alpha u_\alpha=-1$.
Notice we have defined the total accretion rate $\dot{M}$ at the location of the anticipated shock position $r_{sh}$, approximated by equation (\ref{eqn:shockrs}) below.
\added{Spherical symmetry is of course highly idealised and we anticipate future work will be carried out
with greater computational resources in three dimensions to account for anisotropic and angular momentum carrying flows.}

With these expressions one can readily derive equilibrium shock profiles extending from
the upstream to the neutron star surface \citep{Chevalier89,Bernal13}. Although we implemented this option for initializing the flow, in practice we find no advantage to doing that - the relaxation times of the two procedures (whether initialized in a free-fall state globally, or in equilibrated shock profiles) are comparable. We do however, use the equilibrium shock position as derived by \citet{Houck91} to secure the outer radius of the computational grid, $r_{out} = 4 r_{sh}$, where $r_{sh}$ is the (approximate) shock radius
\begin{equation}
r_{sh} = 1.6 \times 10^8 \left(\frac{\dot{M}}{M_\odot ~\text{yr}^{-1}}\right)^{-0.4} ~\text{cm} 
\label{eqn:shockrs}
\end{equation}
for a standard neutron star (with mass $1.4 M_{\odot}$ and radius 10 km), a $\gamma=4/3$ relativistic equation of state, and accretion rate $\dot{M}$.

The initial data is thus reduced to a single parameterization of the mass accretion rate $\dot{M}$. When normalized to the Eddington rate, $\dot{{m}} = \dot{M}/\dot{M}_{Edd}$ defines a dimensionless parameter where
\begin{equation}
\dot{M}_{Edd} = \frac{4 \pi G M_{NS} m_p}{c ~\sigma_T} \approx 3.1\times10^{-9} \left( \frac{M_{NS}}{1.4 M_\odot}\right) ~ M_\odot~\text{yr}^{-1} ~.
\end{equation}

The accreting material composition (for both the hydrodynamics and post-processing calculations) is represented by $(X_H,~X_{He},~X_Z)$, where $X_{H}$ is the hydrogen mass fraction, $X_{He}$ is helium, and $X_{Z}$ is the mass fraction of all elements heavier than helium \added{, and because it does not affect our results much, it is arbitrarily} distributed by relative solar abundances. In this work we consider a predominately helium accretor with $(X_H,~X_{He},~X_Z) = (0.01, ~0.98, ~0.01)$ typical of ultra-compact XRB sources \citep{Cumming03} following hydrogen exhaustion. The crust layer atop the neutron star surface (the reconstructed hydrostatic region between grid boundary and accreting material) is composed of $^{56}$Ni.

\subsection{Time Scales}
\label{subsec:timescales}

Defining a dynamical time $\tau_{dyn}$ as the free-fall time associated with the equilibrium shock radius
\begin{equation}
\tau_{dyn} = \frac{r_{sh}}{v_{ff}} = \frac{r_{sh}^{3/2}}{\sqrt{2 G M_{NS}}} \propto \dot{M}^{-0.6} ~,
\end{equation}
we observe an inverse correlation between $\tau_{dyn}$ and accretion rate $\dot{M}$.
Across the five decades of accretion rates considered in this report, we expect on this scaling alone for the lowest rate calculations to require more than $10^3$ times the computational resources compared to the highest. Of course if the calculations were all run with the same cell resolution, there is the extra cost of having to add more zones for lower accretion rate calculations because of the dependency of the shock radius on $\dot{M}$ (which increases with decreasing $\dot{M}$). This is in fact one of the main reasons we limit our parameter space to $\dot{M} \ge 0.3$ $M_\odot$ yr$^{-1}$.

We can additionally estimate the neutrino cooling time $\tau_\nu$ by the ratio of energy gained to the neutrino loss rate. Assuming uniform accretion and relating the increase of kinetic energy to the potential energy in the free-fall limit (neglecting nuclear burn), we write over a characteristic time interval $\Delta t$
\begin{equation}
\tau_\nu \approx \frac{E_\nu}{L_\nu}
         \approx \frac{G M_{NS}}{R_{NS} L_\nu} \dot{M}\Delta t
         \approx \frac{G M_{NS}}{R_{NS} L_\nu} \rho_\nu 4\pi R^2_{NS} \delta r_\nu ~,
\label{eqn:coolingTime}
\end{equation}
where $L_\nu$ is the neutrino luminosity, $\rho_\nu$ and $\delta r_\nu$ are the mean density
and width of the neutrino cooling layer, and $\Delta t$ corresponds to roughly
when the outward moving shock equilibrates or begins to fall-back onto the NS surface.
If we further link this time interval to a multiple of the dynamical time, the cooling time scales as $\dot{M}^{0.4}/L_\nu \propto  \dot{M}^{-0.85}$, where in the latter expression we used the pair annihilation approximation $L_\nu \propto \dot{M}^{5/4}$ from \citet{Fryer96}.
Although this approximation is understandably crude due to nonlinear interactions between hydrodynamics, nuclear reactions, and neutrino cooling, it nonetheless predicts an inverse scaling with the mass accretion rate close to but somewhat less than linear, a prediction validated by our calculations.

\subsection{Grid and Model Parameters}
\label{subsec:modelParameters}

Table \ref{tab:runs} lists parameters for all 1D and 2D runs, including
accretion rates ($\dot{m}$, $\dot{M}$), maximum spatial resolution at the NS surface ($\Delta r_{min}$), and the estimated (via equation (\ref{eqn:shockrs})) equilibrium shock radius ($r_{rsh}$). Essentially we explore two parameters: the mass accretion rate and spatial resolution covering the neutrino cooling layer for convergence testing.
The run labels identify parameter values, e.g., R08 (R10) represents an accretion rate of $10^8$ ($10^{10}$) Eddington. All calculations were run at both $\Delta r_{min}=0.05$ and $0.1$ km resolutions to assess convergence.

\begin{deluxetable}{lccccc}
\tablecaption{Run Parameters \label{tab:runs}}
\tablewidth{0pt}
\tablehead{
\colhead{\text{run}}               & 
\colhead{$\dot{m}$}           & 
\colhead{$\dot{M}$}                & 
\colhead{$\Delta r_{min}$}         & 
\colhead{$r_{sh}$}                 \\
\text{ }                   & 
($\dot{M}_{Edd}$)          & 
($M_\odot \text{yr}^{-1}$) & 
(\text{km})                & 
(\text{km})                  
}
\startdata
R08  & $10^{8}$   & 0.3   & (0.05, 0.1)  & $2.6\times10^{3}$    \\
R09  & $10^{9}$   & 3     & (0.05, 0.1)  & $1.0\times10^{3}$    \\
R10  & $10^{10}$  & 30    & (0.05, 0.1)  & $4.0\times10^{2}$    \\
R11  & $10^{11}$  & 300   & (0.05, 0.1)  & $1.6\times10^{2}$    \\
R12  & $10^{12}$  & 3000  & (0.05, 0.1)  & $6.4\times10^{1}$    \\
R13  & $10^{13}$  & 30000 & (0.05, 0.1)  & $2.6\times10^{1}$    \\
\enddata
\end{deluxetable}

Figure 1 and Table 1 of \citet{Houck91} indicate that the radius at which cooling and flow times equate is on the order of a kilometer. This suggests a grid resolution of 0.1 km or better is required to model properly the cooling layer, motivating our choices for the minimum cell size. We note that the theoretical shock position (predicted by equation (\ref{eqn:shockrs})) comes to within a factor of two to three of the numerically calculated result (see Table \ref{tab:results_1d}), and over-estimates the calculated value in every case. Hence it is a reasonably good estimate, and at the same time provides a safe measure for setting the outer grid boundary.

The simulations are run on a spherical grid using a logarithmic radial coordinate to achieve high resolution near the NS surface where neutrino production is important, while relaxing resolution exponentially near the outer boundary according to
\begin{equation}
r = r_0 e^{(\eta-1)/\eta_B} ~.
\end{equation}
Here $\eta$ is the logarithmic coordinate, $r_0 = (1 + \delta \Delta r_{min}) R_{NS}$ defines the inner radius of the grid, $R_{NS}$ is the neutron star radius, $\delta = 10^{-2}$ is an arbitrary small number, $\eta_B=2$ controls the logarithmic spacing, and the number of radial zones is determined by
\begin{equation}
N_r = \eta_B \log \left( \frac{r_{out}}{r_{0}}\right) \left( \frac{r_{0}}{\eta_B ~\Delta r_{min}} \right) ~.
\end{equation}
The number of zones thus depends on both the grid resolution and accretion rate 
(which sets the radius of the outer boundary $r_{out}$ as a multiple of $r_{sh}$). 
At standard resolution ($\Delta r_{min}=0.1$ km) $N_r$
varies from 320 to 700 (highest to lowest accretion rate), and twice that
for the high resolution cases  ($\Delta r_{min}=0.05$ km).
The 2D calculations are run only at the standard radial resolution
while maintaining a fixed number of angular zones $N_\theta=400$.

\section{1D Results}
\label{sec:1Dresults}

\begin{figure}
\hspace{0.0in}\includegraphics[width=1.1\textwidth]{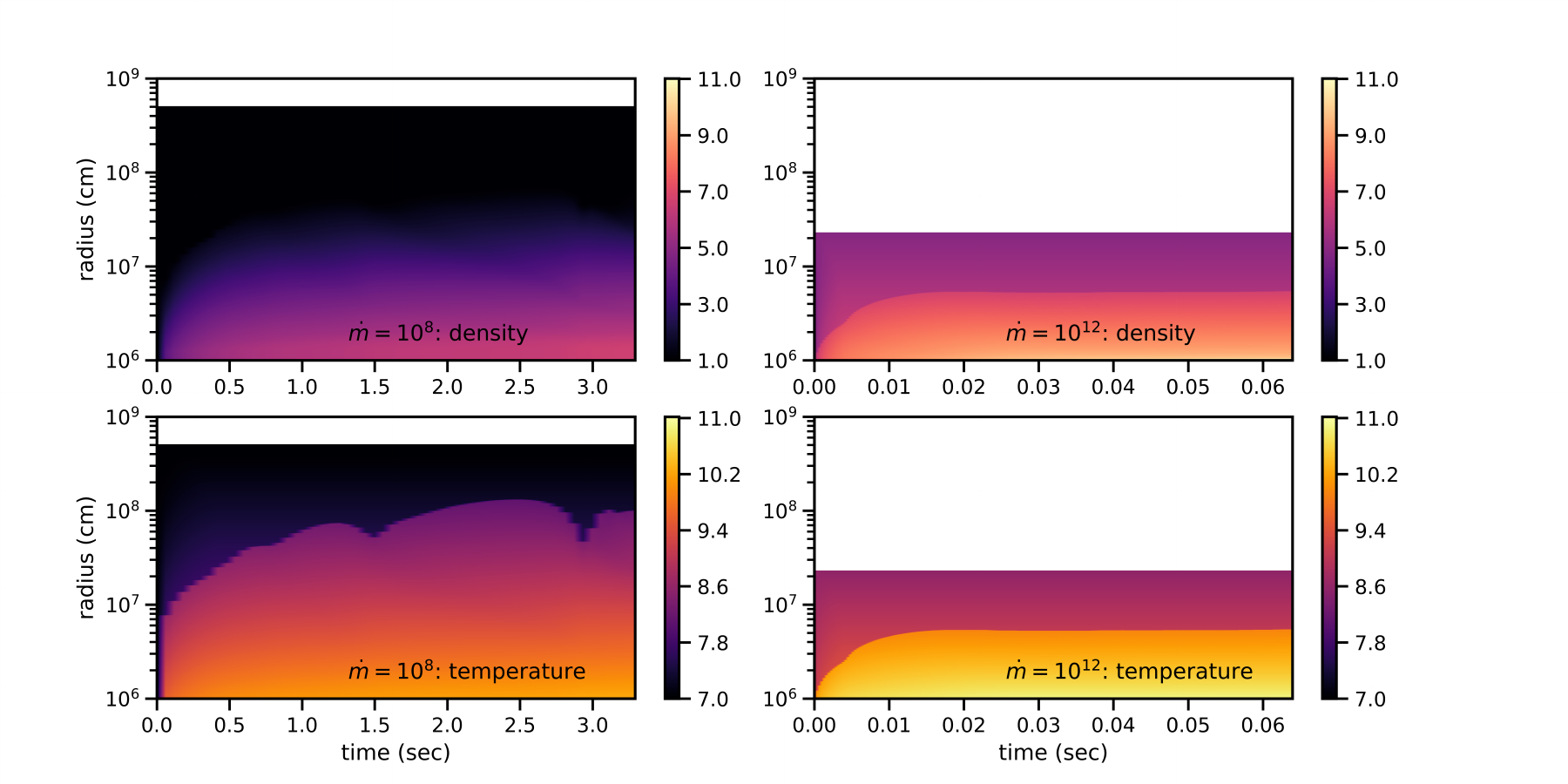}
\caption{
Spacetime images showing the development and propagation of the shock generated
by hyper-accreting material entering the grid from the top and reflecting off
the NS surface at the bottom. The images plot the logarithms of density \added{(g/cm$^3$)} and temperature \added{(Kelvin)} as a
function of time from low ($\dot{{m}} = 10^{8}$)
and high ($\dot{{m}} = 10^{12}$) accretion rate models.
}
\label{fig:contours}
\end{figure}

We begin with studies of the one-dimensional shock structure, including effects
of nuclear burn and neutrino cooling. Figure \ref{fig:contours} plots spacetime (radius versus time) images of density and temperature from two representative cases: $\dot{{m}} = 10^{8}$ and $\dot{{m}} = 10^{12}$. As material accretes onto the neutron star from large radii, the resulting shock propagates outward (upward in the images) from the star surface. Over time, as the post-shock region gets denser and hotter, neutrino cooling activates strongly in a very thin layer near the NS surface. Eventually material cools enough
to affect the velocity of the outward moving shock, causing the shock to equilibrate
or even periodically fall back towards the NS as in case R08. 

\begin{figure}
\hspace{1.2in}\includegraphics[width=0.6\textwidth]{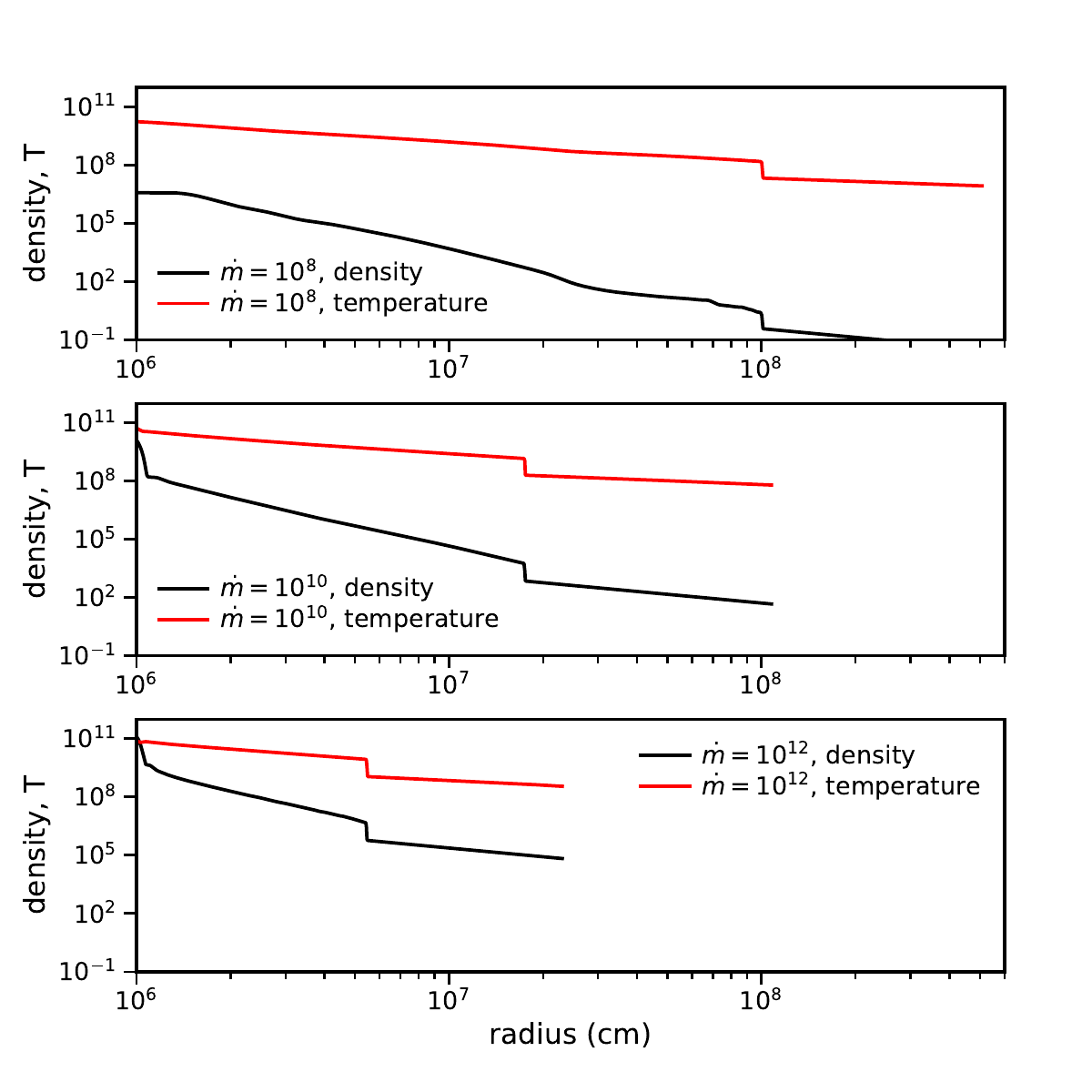}
\caption{
Late time line profiles of density (black) and temperature (red) from the
$\dot{{m}} = 10^{8}$, $\dot{{m}} = 10^{10}$, and $\dot{{m}} = 10^{12}$
cases.
}
\label{fig:rhoT}
\end{figure}

Line profiles of the density and temperature for low, intermediate, and high
accretion rate models are plotted in Figure \ref{fig:rhoT} at the final times
corresponding roughly to when the outgoing shocks in Figure \ref{fig:contours} reach maximum extent.
The line profiles are characterized by three distinct regions: the pre-shocked matter falling in from the right and
compressing adiabatically due to spherical convergence; the post-shocked gas that continues to
fall towards the neutron star with a somewhat steeper profile; and the thin neutrino cooling
layer near the neutron star surface that sharply increases the density profile well above
the equilibrium solution.

Hydrodynamic and cooling diagnostics are summarized in Table \ref{tab:results_1d} where we tabulate the
settled time-averaged accretion rate one kilometer from the neutron star surface $\dot{m}_S$,
the dynamical time $\tau_{dyn}$, cooling time $\tau_{cool}$, time for the shock to reach maximum extent $\tau_{rsh}$,
neutrino cooling layer thickness $\delta r_{\nu}$, maximum radial extent of the shock $r_{sh}$ (which can
be compared to Chevalier's approximation in Table \ref{tab:runs}), and the neutrino luminosity $L_{\nu}$ 
and nuclear energy release rate $\dot{e}_{nuc}$ at the end of each calculation. $\delta r_{\nu}$ is estimated from the size of the post-shock region near the NS surface where the density increases sharply due to recompression from neutrino energy loss. We note that this feature is barely discernable \added{due to the lack of neutrino cooling} at the lowest accretion rate $\dot{{m}} = 10^{8}$ (see Figure \ref{fig:rhoT}), so we leave those entries blank in the table. \deleted{ but assign it (arbitrarily) the R09 value in equation (\ref{eqn:coolingTime}) to demonstrate the orders of magnitude differences in the R08 cooling rate from the others.}

\begin{deluxetable}{lccccccccc}
\tablecaption{1D Result Summary \label{tab:results_1d}}
\tablewidth{0pt}
\tablehead{
\colhead{\text{run}}        & 
\colhead{$\dot{{m}}_S$} & 
\colhead{$\tau_{dyn}$}      & 
\colhead{$\tau_{cool}$}     & 
\colhead{$\tau_{rsh}$}      & 
\colhead{$\delta r_{\nu}$} & 
\colhead{$r_{sh}$}          & 
\colhead{$L_{\nu}$}         &
\colhead{$\dot{e}_{nuc}$}   \\
\text{ }                    & 
($\dot{M}_{Edd}$)          & 
(sec)          & 
(sec)          & 
(sec)          & 
(km)           & 
(km)           & 
(erg/s)        & 
(erg/s)        & 
}
\startdata
R08  & $2\times10^{7}$   & 0.6    & -      &  2.5    & -    & $1.3\times10^3$  & $1.2\times10^{45}$  & $2.2\times10^{44}$   \\
R09  & $3\times10^{9}$   & 0.15   & 0.003  &  1.5    & 1.3  & $6.0\times10^2$  & $4.4\times10^{46}$  & $4.3\times10^{45}$   \\
R10  & $5\times10^{10}$  & 0.06   & 0.002  &  0.35   & 1.0  & $2.3\times10^2$  & $3.6\times10^{48}$  & $1.8\times10^{47}$   \\
R11  & $7\times10^{11}$  & 0.02   & 0.002  &  0.07   & 0.9  & $1.1\times10^2$  & $4.7\times10^{49}$  & $1.3\times10^{48}$   \\
R12  & $4\times10^{12}$  & 0.006  & 0.001  &  0.02   & 0.9  & $5.7\times10^1$  & $4.1\times10^{50}$  & $5.1\times10^{48}$   \\
R13  & $4\times10^{13}$  & 0.001  & 0.001  &  0.005  & 0.9  & $3.3\times10^1$  & $3.4\times10^{51}$  & $1.0\times10^{50}$   
\enddata
\end{deluxetable}

\begin{figure}
\hspace{1.0in} \includegraphics[width=0.6\textwidth]{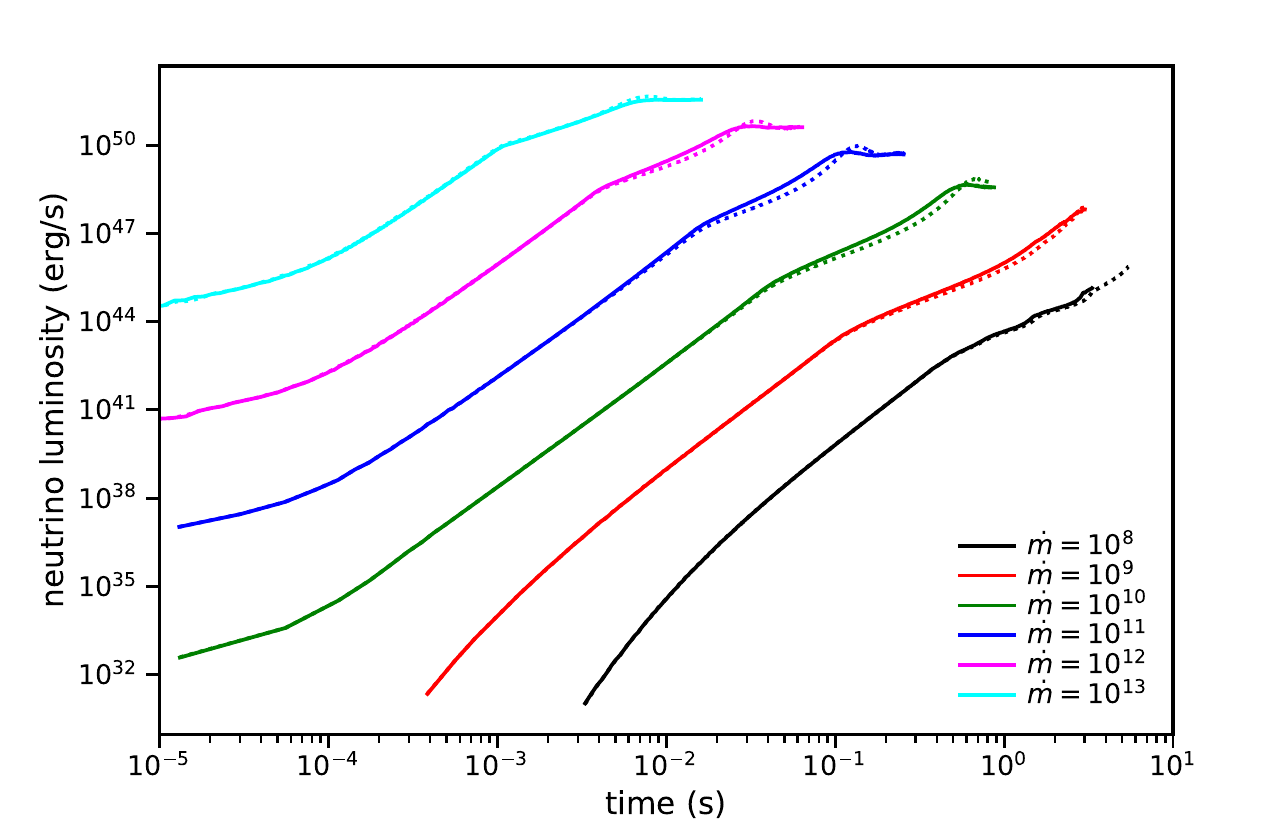}
\vspace{0.2in}\caption{
Neutrino luminosity is plotted as a function of time for all accretion models.
Solid (dotted) lines represent results from calculations run with (without) nuclear burn.
}
\label{fig:LvsTime}
\vspace{0.2in}
\end{figure}

\begin{figure}
\hspace{1.0in} \includegraphics[width=0.6\textwidth]{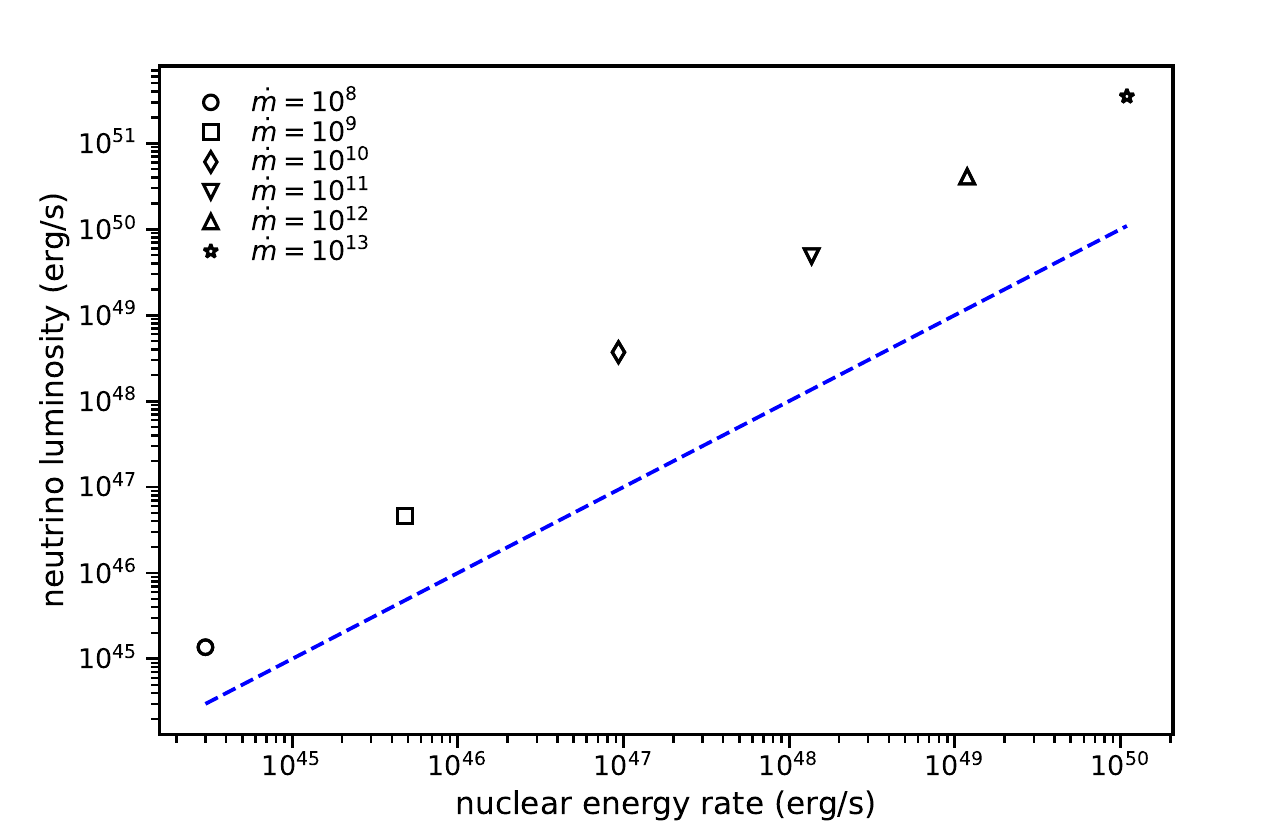}
\caption{
Neutrino luminosity is plotted versus nuclear energy rate at a time
when the shock reaches maximum extent.
The dashed line is a reference corresponding to $\L_\nu = \dot{e}_{nuc}$.
}
\label{fig:LnuvsLnuc}
\end{figure}

Cooling (and heating) rates are plotted in Figures \ref{fig:LvsTime} and \ref{fig:LnuvsLnuc}.
Figure \ref{fig:LvsTime} shows the neutrino loss rate (luminosity) in units of ergs/s as a function of time
for each of the cases. Notice case R13 develops a neutrino luminosity typically associated with core collapse \citep{Pejcha12}. Figure \ref{fig:LvsTime} additionally compares energy rates
from calculations that included nuclear burn (solid lines) to calculations where burn was deactivated (dotted lines). Apart from slight differences at late times from the higher accretion rate calculations, these results suggest feedback effects from nuclear burn do not significantly impact neutrino emissions nor the hydrodynamic progression of the accretion shock, resulting in only a modest (less than 30\%) change in the equilibrium shock positions.
This conclusion is confirmed also by Figure \ref{fig:LnuvsLnuc} which plots neutrino luminosity against the nuclear energy rate. Depending on the accretion rate and the time this comparison is made, we find neutrino cooling dominates over nuclear energy production (or absorption) by at least an order of magnitude. The data plotted in Figure \ref{fig:LnuvsLnuc}
correspond roughly to when the shock first reaches maximum extent.

\begin{figure}
\hspace{1.0in}\includegraphics[width=0.7\textwidth]{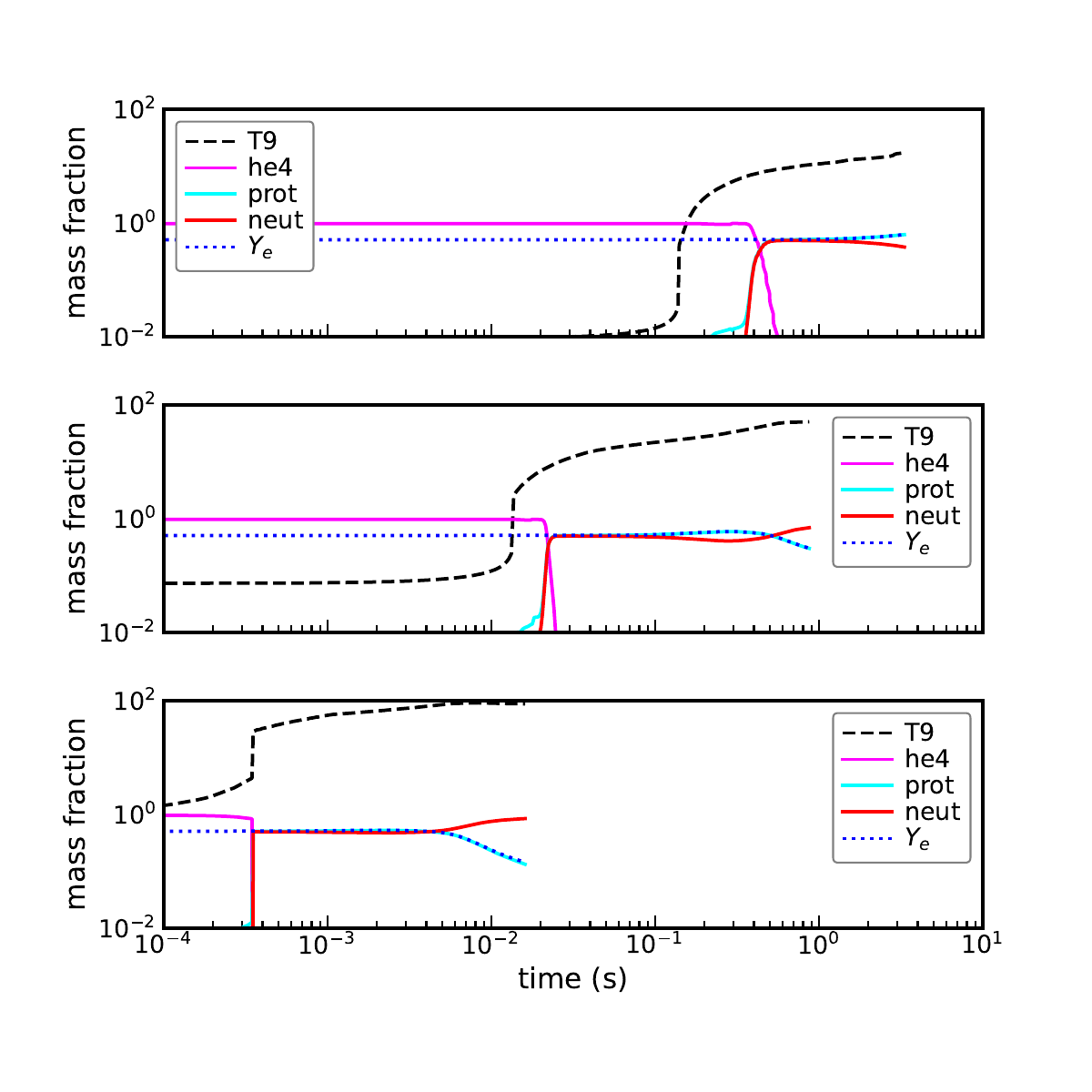}
\caption{
Time histories of the electron fraction $Y_e$, and the neutron, proton, and helium mass fractions for the $\dot{{m}} = 10^{8}$ (top), $10^{10}$ (middle) and $10^{13}$ (bottom) calculations. Also shown is the temperature (black dashed lines) in units of $10^9$ K. Solutions are derived by post-processing Lagrangian tracers with Winnet and a 7150 element reaction network, following their evolution through adiabatic compression and shock heating before they enter the cooling layer and settle onto the NS surface. 
}
\label{fig:speciestrack}
\end{figure}

Although energy changes (sources or sinks) from nuclear activity do not affect the accretion shock much, nucleosynthesis nevertheless determines the development of neutron versus proton-rich sites. This is demonstrated by Figure \ref{fig:speciestrack} which plots time histories experienced by a typical particle tracer from the $\dot{{m}} = 10^{8}$, $10^{10}$, and $10^{13}$ models. These results are derived by post-processing trajectories with a more detailed calculation using the Winnet nuclear reaction code and a 7150-isotope network in place of the inlined 19-isotope model. Each particle track begins at a radius of $r=r_{sh}$, accreting towards the neutron star as a Lagrangian fluid tracer, before crossing the shock to a new adiabat and falling into the dense cooling layer along the neutron star surface where particles stagnate due to hydrostatic boundary conditions. We note however that particle tracers are expected to
behave differently in multi-dimensions due to convective instabilities: they will not remain at the NS surface, but instead get caught up in convection currents and therefore experience repeated cycles of cooling and heating as they rise and fall between the neutron star surface and the shock. It is nevertheless interesting to point out that the isotopic distributions in these 1D calculations are initially dominated by helium prior to shock passage, but in all cases the post-shock conditions are extreme enough to dissociate elements almost entirely into neutrons and protons. This occurs immediately at shock passage in the highest accretion rate models, but is systematically delayed at lower rates due to lower densities and temperatures. Of particular interest is the development of both proton and neutron-rich compositions when tracers approach the cooling layer. The former is driven primarily by positron captures at relatively low accretion rates (top plate of Figure \ref{fig:speciestrack}); the latter by electron capture reactions which dominate at the higher densities observed in the highest accretion rate models (bottom two plates of Figure \ref{fig:speciestrack} and Figure \ref{fig:rhoT}).

\section{2D Results}
\label{sec:2Dresults}

\subsection{Convective instability}
\label{subsec:convection}

In spherical symmetry, tracers fall to the NS surface exhibiting little variability other than correlating electron fraction to accretion rate, trending towards neutron richness with increasing $\dot{m}$ (Figure \ref{fig:speciestrack}). When spherical symmetry is relaxed, the entropy graded atmosphere drives convective instabilities which develop currents capable of cycling fluid between the star surface and outgoing shock. Turbulence drives the shock well beyond the spherical equilibrium radius, while circulating fluid at velocities small compared to the escape velocity ($v_{esc} \approx 1.9\times10^{10}$ cm/s) but with high enough specific energies (equation (\ref{eqn:unbound})) to become unbound from the neutron star.

\begin{figure}
\hspace{0.7in}\includegraphics[width=0.9\textwidth]{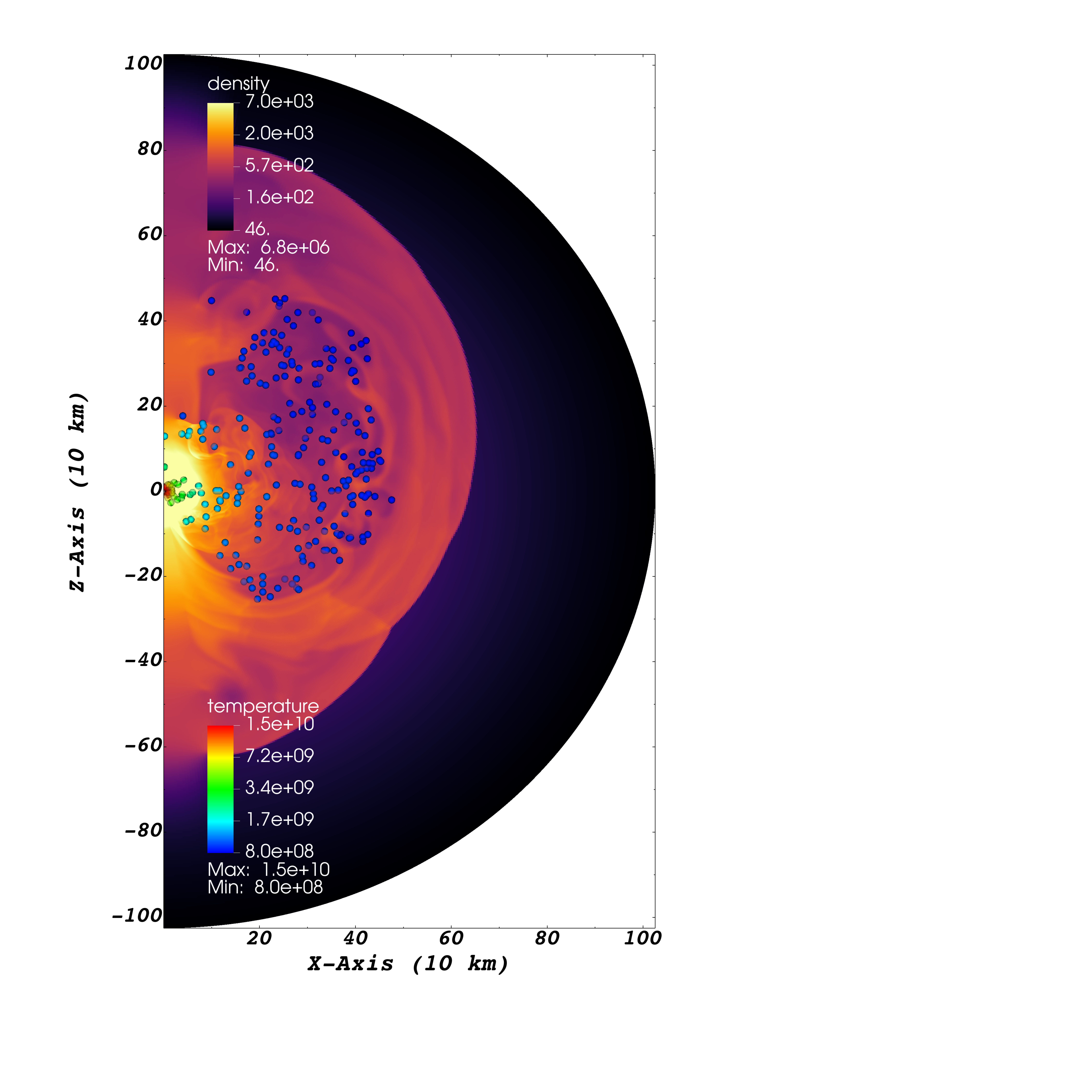}
\caption{
The convective instability demonstrated by
the logarithm of gas density (gm/cm$^3$) for model $\dot{{m}}=10^{10}$
at the final computed time, 0.087 seconds. Also shown are tracer particles
from the inner-most ring of tracers, color coded to the gas temperature (degrees Kelvin).
Some tracers (with temperatures in excess of 7 GK) have fallen into the neutrino sphere,
but many are caught in convection cells and cycle between the neutrino sphere and hydrodynamic shock.
}
\label{fig:convection}
\end{figure}

Figure \ref{fig:convection} shows these convection currents by plotting the logarithm of the gas density at the last evolved time (0.087 seconds) for the $\dot{{m}}=10^{10}$ case. Also shown are a representative sample of tracer particles colored according to the gas temperature. Notice the multi-scale nature of these convective cells common to turbulent flows, with numerous small local cells operating within larger super-structures extending out to $10^8$ cm dimensions. Also notice the broad distribution of particle positions, with a significant number of them following the currents, while others fall into the neutrino sphere or are captured by the neutron star 
(in which case they are dropped from the simulation and don't appear in the image).
All of our simulations develop similar structures. The primary differences affecting nucleosynthesis are the post-shock densities and temperatures (e.g., Figure \ref{fig:rhoT}), and by association their dimensionless entropy
\begin{equation}
\widetilde{S}_r \equiv \frac{S_r}{k~N_A} \approx 5.2 ~\left(\frac{T}{1.16\times10^{10}}\right)^3 \left(\frac{\rho}{10^8}\right)^{-1} ~,
\end{equation}
accounting for radiation and electron-positron pairs \citep{Woosley92}.

\added{To experience different instances of shock crossings and post-shock environments, we initialize four spherical rings of tracers, each consisting of 250 particles spaced uniformly in radius between the outer grid boundary and a fraction of the equilibrium shock radius. They are also distributed about the polar direction to evenly sample the enclosed volume, and because the initial data is spherically symmetric, the tracers in each ring effectively represent equal mass elements.} \added{Tracers are prevented from falling onto the NS before the shock fully develops,} by holding them in place (prevented from following the fluid) until the temperature exceeds $5\times10^8$ K, after which they are released to track fluid elements in Lagrangian fashion as they shock heat and either fall to the neutron star surface (in which case they are dropped from the tally), or get swept up by convection cells. Along the way, tracers record (and store) local densities, temperatures, velocities, and radius as a function of time. The accumulated data is then filtered to meet the following three criteria before being post-processed for nuclear yields: (1) they must have met a threshold maximum temperature of 7 GK for potential disassembly, (2) they must have cooled below 5 GK by the end of their trajectory to have initiated re-assembly, and (3) they must have acquired sufficiently high specific energies and velocities to satisfy the following unbound condition:
\begin{equation}
u_0 = g_{0\alpha} u^\alpha = 
      g_{0\alpha} u^0 v^\alpha \approx -\left( \epsilon + v^i v_i \right) \left(\frac{r}{2 G M_{NS}}\right) < -1 ~,
\label{eqn:unbound}
\end{equation}
where $g_{\alpha\beta}$ is the spacetime metric, $u^\alpha$ the four velocity, $v^i$ the spatial tracer velocity,
and $\epsilon$ the specific internal energy.

We find that none of the particle trajectories from the two lowest accretion rate 
calculations ($\dot{{m}}=10^{8}$ and $10^{9}$)
generate sufficiently hot post-shock temperatures to meet the first filter criterion.
We additionally find the highest rate calculation ($\dot{{m}}=10^{13}$) extremely sensitive to numerical
resolution, and computationally expensive due to increased nuclear activity (which dominates
computational time).  Because post-shock temperatures are particularly high in this case, we have not
been able to run this simulation late enough to have cooled tracers below the second criterion.
Finally, and perhaps most interestingly, we observe an anti-correlation between the mass range of nuclear
end-products and accretion rate among the three cases which meet all three criteria, $\dot{{m}}=10^{10} - 10^{12}$.
We therefore restrict our post-processing analysis to the $\dot{{m}}=10^{10}$ model which makes the heaviest elements, and leave a more complete analysis and comparison against the other models to future work.

\begin{figure}
\hspace{0.5in}\includegraphics[width=0.8\textwidth]{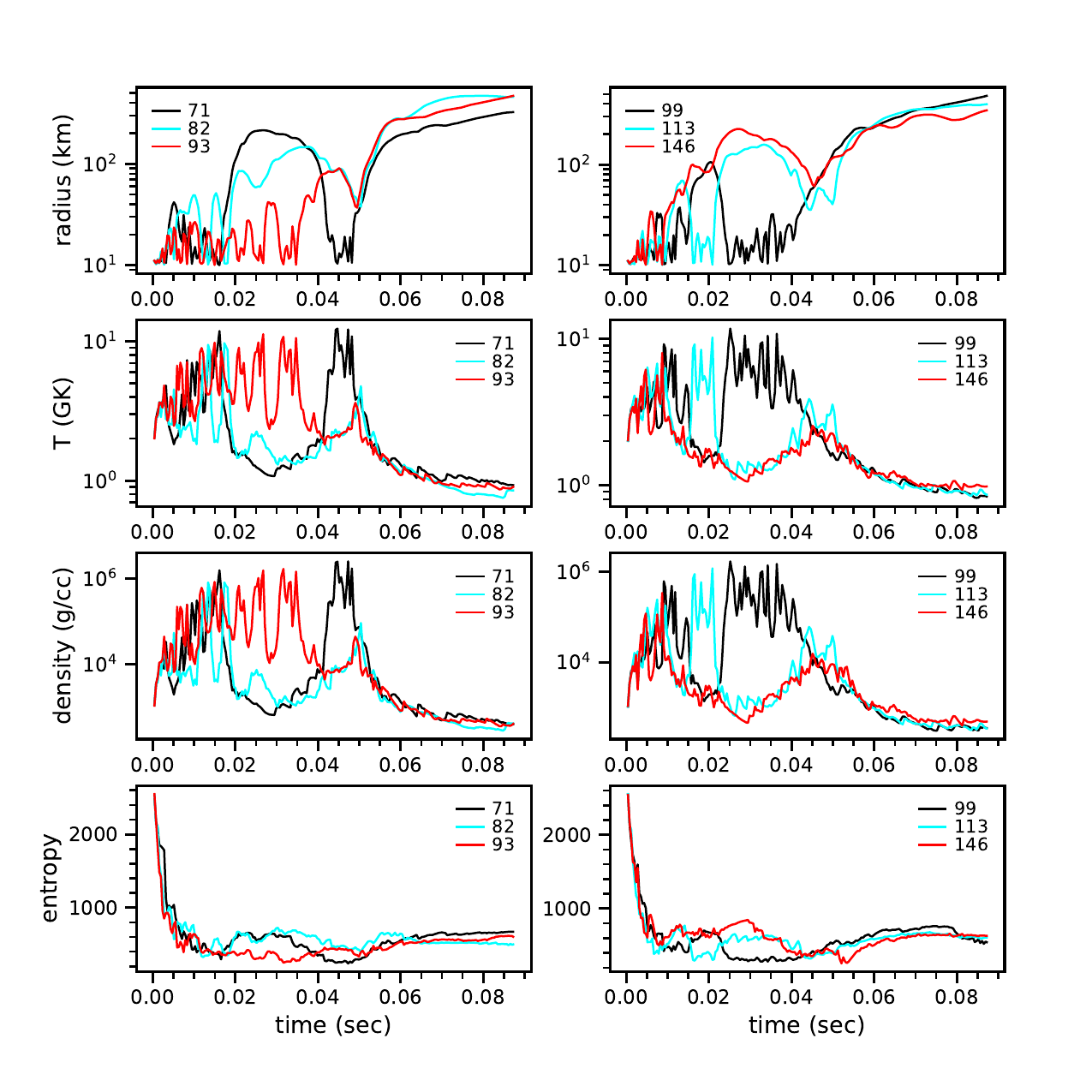}
\caption{
Time histories of typical particle tracks extracted from the $\dot{{m}} = 10^{10}$ calculation 
showing the trajectory radius (top row), temperature (second row),  gas density (third row), and
dimensionless entropy (bottom row).
}
\label{fig:massfractionvstime}
\end{figure}

From the $\dot{{m}}=10^{10}$ model, we focus on the following set of tracks representing a broad range of nuclear yield outcomes: 82, 113, 93, 146, 99, and 71, ordered from lowest atomic mass elements produced to highest. 
To get a sense of the cyclic time scales and thermal environments experienced by 
these tracers, we plot in Figure \ref{fig:massfractionvstime} the radius (top row), temperature (second row),
density (third row), and entropy (bottom row) as a function of time for several tracers.
Notice in the early stages, when the shock is still relatively close to the neutron star, the average
periodicity or turnover time scale is relatively short (a couple milliseconds), but increases
to about 10 milliseconds as the shock moves away from the neutron star. Eventually, by $\sim0.06$ seconds  trajectories break the cyclic pattern and settle onto outward-directed radial tracks, moving at nearly 
constant velocities between 0.01 - 0.03$c$.

The average circulation or turnover period can be estimated from the top plate of Figure \ref{fig:massfractionvstime}
or by averaging tracer velocities across multiple convection periods $\tau_{conv} \approx r_{sh}/\overline{v}$. Either way we find $\tau_{conv}$ to be no more than a fraction of a second. This timescale can be compared to the
approximate time required for the neutron star to collapse to a black hole from accreting material \citep{Fryer96}
\begin{equation}
\tau_{BH} = \frac{1}{L_\nu} \frac{GM_{NS} \delta m}{r_{NS}} \approx 10^4 ~\text{secs}  ~,
\end{equation}
where $L_\nu$ is the total neutrino emission per unit time (from Table \ref{tab:results_1d}), 
$\delta m$ is the additional mass
required to induce collapse (assuming 0.1 $M_\odot$), and $r_{NS}$ is the neutron star radius.
Clearly $\tau_{conv} \ll \tau_{BH}$, suggesting an equilibrium atmosphere is likely to develop
in this model long before the neutron star collapses to a black hole.

\subsection{Nuclear yields}
\label{subsec:yields}

\begin{figure}
\hspace{0.1in}\includegraphics[width=1.0\textwidth]{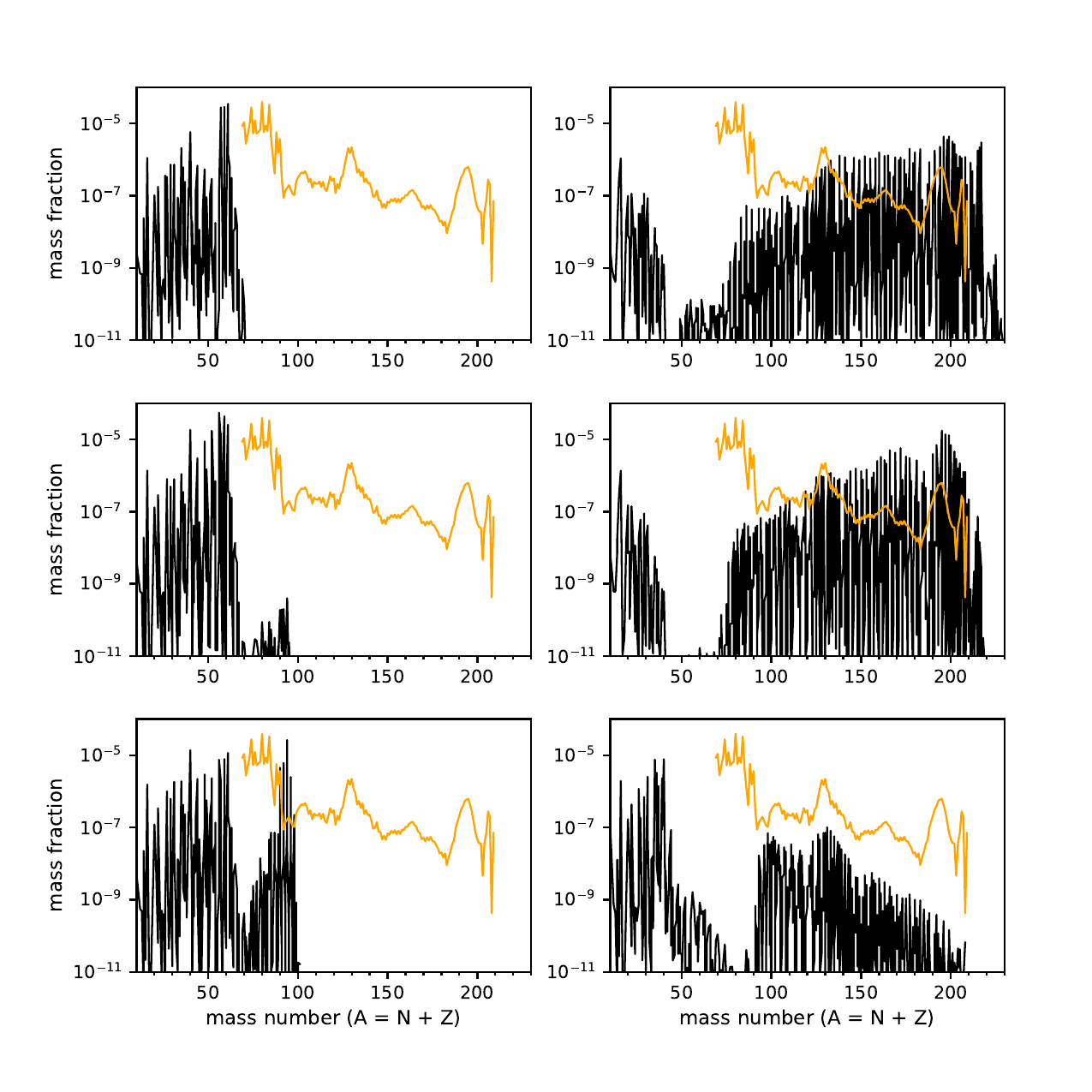}
\caption{
Final mass fractions observed by post-processing six separate tracer histories from the $\dot{m}=10^{10}$ model, demonstrating a broad diversity in mass distributions. The black lines are results calculated by the Winnet reaction code with a 7150 element network. Orange curves are solar system abundances \citep{Goriely:1999} normalized to the iron mass fraction.
}
\label{fig:n3_tracers}
\end{figure}

Nuclear activity is calculated  by post-processing thermodynamic tracer histories with the Winnet reaction code and a 7150 isotope network consisting of elements through Rg$^{339}_{111}$. (Although we do not show the equivalent outputs from PRISM, we have confirmed they are nearly identical to Winnet's and XNet's.) Figure \ref{fig:n3_tracers} plots the final mass fraction distributions from the six representative tracks (identified in the previous section) as a function of atomic mass. The mass fractions correspond to the end times of the simulations after the tracks break from their cyclical trajectories and transition to an adiabatically expanding or cooling phase with essentially locked-in distributions. 

\begin{figure}
\hspace{0.8in}\includegraphics[width=0.6\textwidth]{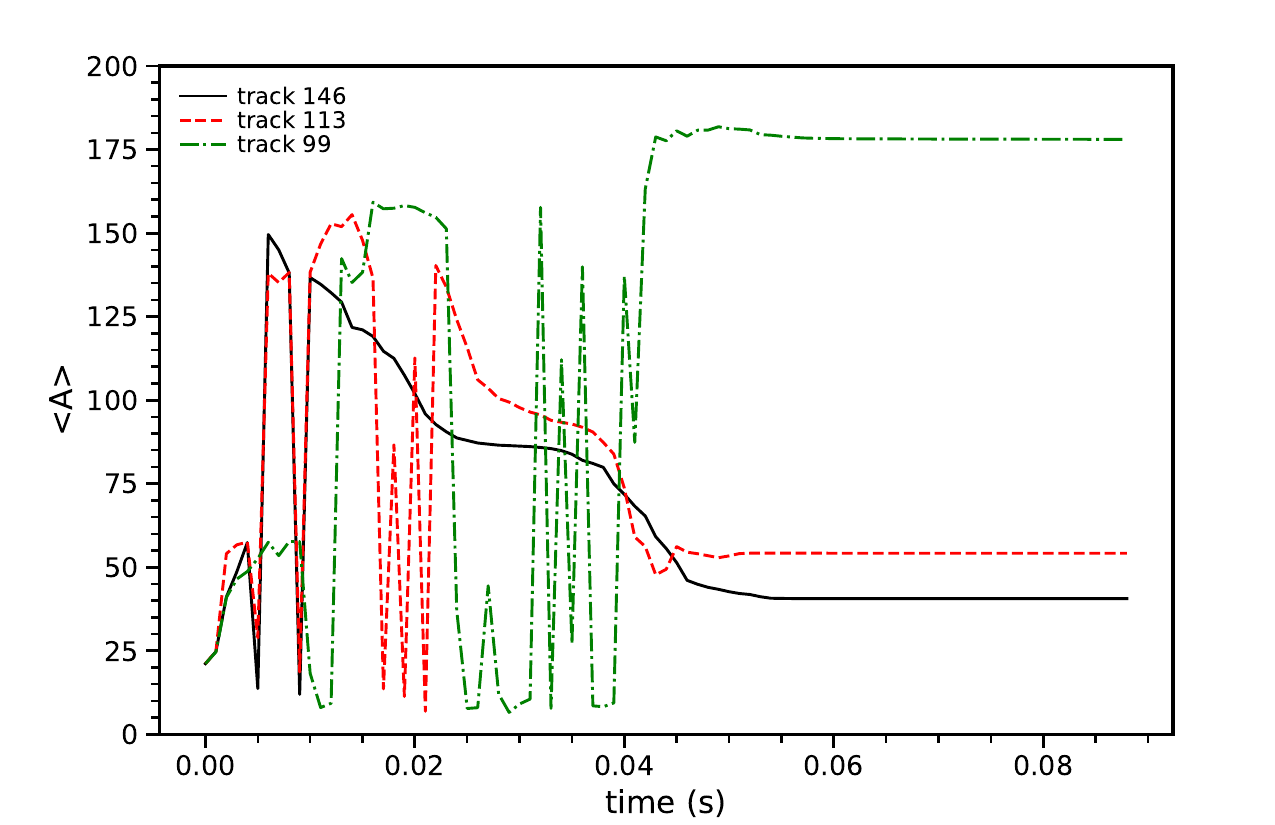}
\caption{
Average atomic masses weighted by mass fractions and plotted as a function of time. 
Because of its dominance we have ignored helium in these calculations to get a more enlightened representation of the mass distribution.
}
\label{fig:avgA}
\end{figure}

The steady-state nature of the final nucleon distributions is demonstrated in Figure \ref{fig:avgA} where we plot the average atomic mass, weighted by mass fractions and neglecting helium, as a function of time for a few tracks. Up to about 60 milliseconds, all tracers experience multiple episodes of photodisintegration and heavy element synthesis as they cycle between the shock and neutrino sphere. Each time the temperature increases, the composition resets to mostly protons, neutrons and $\alpha$ particles. At high temperatures ($>8$~GK) the calculations evolve in nuclear statistical equilibrium, but we also solve for weak reactions which may affect the electron fraction. However, we find that neither neutrino interactions nor electron captures affect the electron fraction in any noticeable way. The neutrino flux estimated at tracer locations (assuming supernova-like neutrino spectra) does not exceed $10^{40}$cm$^{-2}$, which is orders of magnitude less than the fluxes expected from neutrino-driven winds \citep{Qian96}. Additionally, the density typically does not exceed $10^6$~g/cm$^3$ (see Figure \ref{fig:massfractionvstime}), too low for robust electron capture reactions. \replaced{After 60 milliseconds, the mean atomic mass flattens as the mass distribution freezes out during the exit cycle, and it is this last cooling phase which shapes the final composition.} {After several cycles of photodisintegration and recombination of nuclei, the tracer particles enter their final phase of expansion, after which complete photodisintegration does not occur again. This final freeze-out phase establishes the composition of the ejected material and it can itself be non-monotonic, as discussed in detail below. The nucleosynthesis during this final exit cycle ends typically after 60 milliseconds, when the mean atomic mass flattens as the mass distribution freezes out.}

\begin{figure}
\hspace{0.8in}\includegraphics[width=0.7\textwidth]{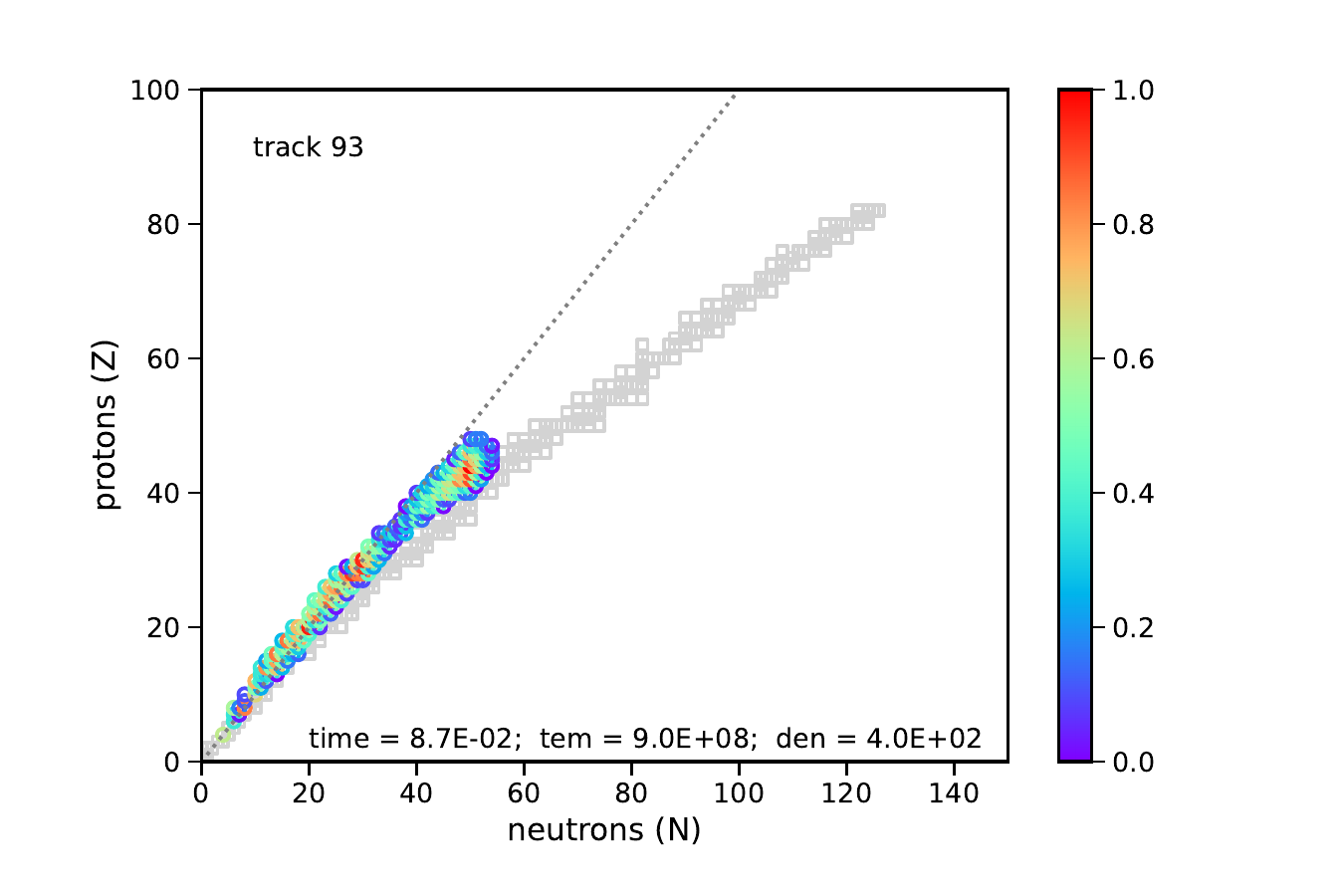}
\caption{
The final distribution of nuclei from track 93 plotted as color-coded circles on the nuclide chart. For clarity, only nuclei with $A>4$ and mass fractions exceeding $10^{-12}$ are shown. The color bar is the logarithm of the relative mass fractions, the dotted line is the equal proton/neutron reference, and grey squares represent stable nuclei.
}
\label{fig:nuchart93}
\end{figure}

Figure \ref{fig:n3_tracers} illustrates the wide range of nuclear yields \replaced{festering}{made} in these hot environments. The left column plates are examples where production stops at $A\leq 100$. For track 82, the final abundance pattern does not extend beyond the iron group, resulting in a final composition virtually unchanged from the NSE composition and dominated by $^{59}$Co and $^{59,60}$Ni. Nucleosynthesis from track 113 extends to heavier isotopes, but the mass fractions for $A>70$ are almost negligibly small. Track 93 shows a bimodal pattern with one peak around the iron group and a second peak around $A$=90, including very large abundances of the $p$-nuclei $^{92,94}$Mo. The proton-rich nature of this particular track is emphasized by Figure \ref{fig:nuchart93}, which displays the final distribution settling above stability on the nuclide chart.

\begin{figure}
\hspace{0.8in}\includegraphics[width=0.7\textwidth]{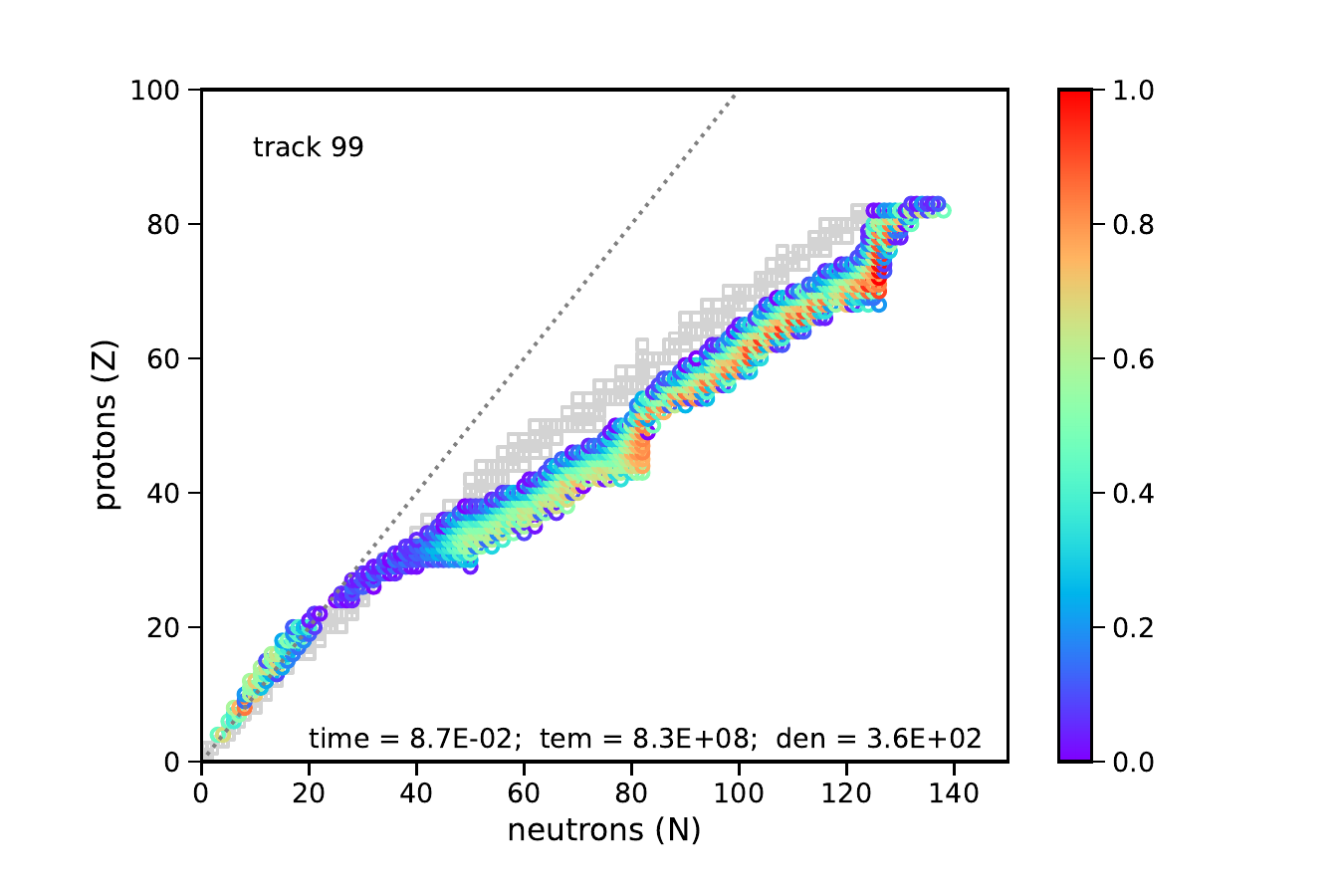}
\caption{
As Figure \ref{fig:nuchart93} except for track 99.
}
\label{fig:nuchart99}
\end{figure}

The right column of panels in Figure \ref{fig:n3_tracers} represent tracks 71, 99 and 146, sampling conditions that allow for the production of heavy elements up to Pb and Bi. Nucleosynthesis in tracks 71 and 99 progressed on the neutron-rich side of stability via the $r$-process, resulting in nuclide distributions represented by Figure \ref{fig:nuchart99} from track 99. The $r$-process in this case reached to the neutron shell closure at $N=126$, but mostly populated isotopes relatively close to stability. These tracks are examples of a high-entropy $r$-process that can occur even without a neutron excess because of a very short expansion timescale, similar to the neutrino-driven winds studied by \cite{Meyer02}. High neutron-to-seed ratios do not require a neutron-rich environment, so that heavy nuclei can form even if $Y_e > 0.5$ due to a persistent disequilibrium between free nucleons and alpha particles at sufficiently high entropy and rapid expansion. Defining the expansion time scale as an e-folding time of the trajectory radius, we find typical expansion times of roughly a few milliseconds. Meyer does not observe the $r$-process at these timescales (only at sub-milliseconds), but the relatively higher entropy of our environments, by roughly a factor of two, allows the $r$-process to proceed at longer freeze-out timescales.

\begin{figure}
\hspace{0.8in}\includegraphics[width=0.7\textwidth]{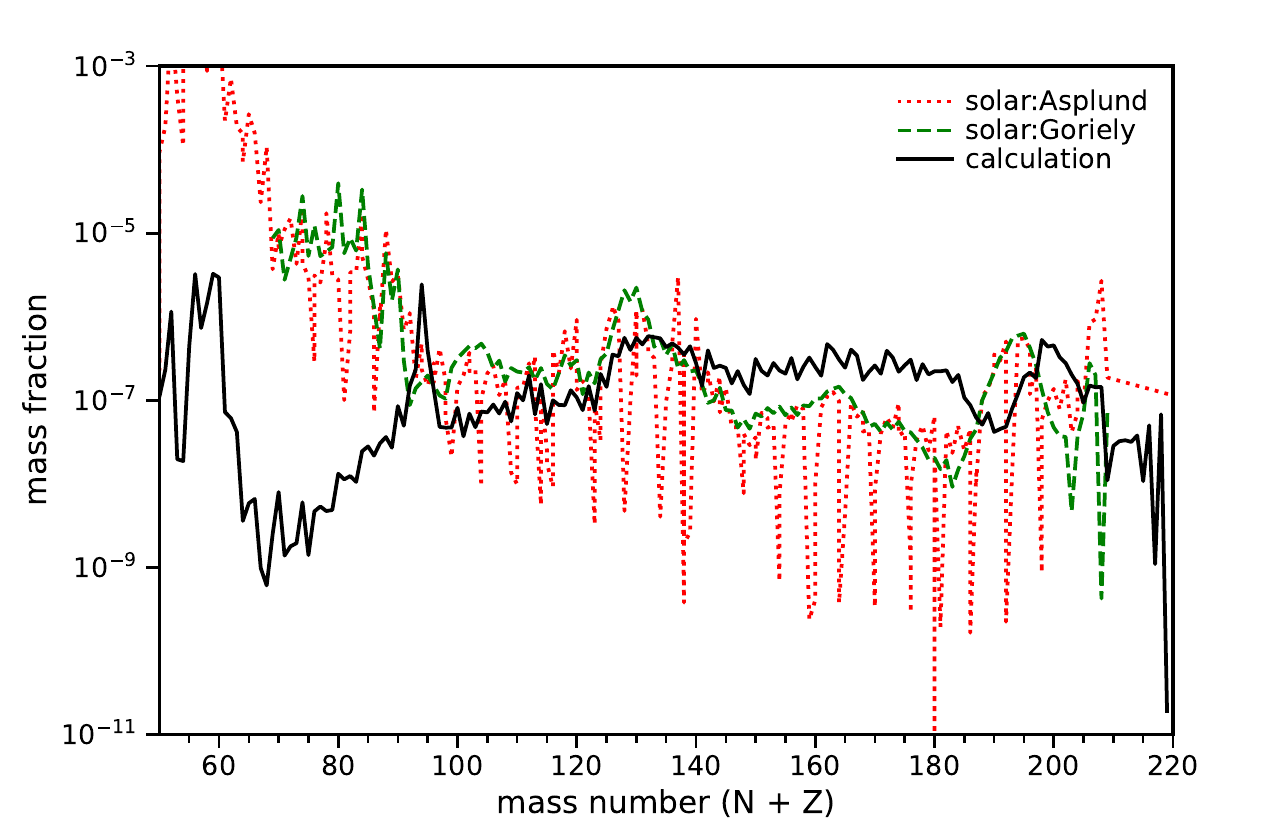}
\caption{
Mass fractions summed over all escaping tracer particles from the $\dot m = 10^{10}$ model, including those that produced $r$-process elements together with those that did not. The colored data represent solar system abundances scaled to the model iron mass fraction.
}
\label{fig:solar_alltrks}
\end{figure}

\citet{Meyer02} additionally pointed out that abundance patterns produced by high-entropy conditions do not match the solar pattern very well. This is also clear in our calculations, as can be seen by comparing Figure \ref{fig:n3_tracers} with the solar abundances plotted in Figure \ref{fig:solar_alltrks}. For example, the final  pattern for track 99 does not show a clear 2nd $r$-process peak around $A=130$. Instead, there is an almost exponential increase of mass fractions from $A=120$ to $A=180$. The third r-process peak forms at around $A=200$, which is between the solar $r$-process peak and the solar s-process peak. In characteristically high-entropy conditions, the neutron density is low and the $r$-process path is closer to stability as confirmed by Figure \ref{fig:nuchart99}. This leads to a different dynamic in which the closed neutron shells are populated at lower $A$, compared to a low-entropy r-process.

Track 146 evolves in a very different manner. Its final abundance pattern contains significant fractions of p-isotopes which cannot be made by neutron-captures. This is the result of a new kind of process, one that to our knowledge has not been previously discussed in the literature. During the first few milliseconds following freeze-out, neutron captures produce elements up to Pb and Bi, while the temperature cools to around 3~GK. Over the next 10 to 20 milliseconds, the neutron-to-seed ratio drops rapidly as neutrons are exhausted and the temperature slowly cools below 2~GK. The abundances remain in a stable (n,$\gamma$)-($\gamma$,n) equilibrium through most of this period. At around 30 milliseconds, however, the temperature of the track increases again above 2~GK (see Figure \ref{fig:massfractionvstime}), which gives rise to a burst of $(\gamma,n)$ and proton-induced reactions that drive the composition from the neutron-rich isotopes of the r-process path toward stability. The high-entropy $(n,\gamma)$-$(\gamma,n)$ equilibrium does not populate isotopes close to the neutron drip-line, but mostly isotopes with half-lives exceeding 100~ms, i.e., $\beta$ decays only play a minor role. 

For the neutron-rich isotopes in track 146 with $Z<40$, $(p,n)$ are the dominant reactions for driving material back to stability during the temperature increase at 30~ms. For heavier elements, the large Coulomb barrier reduces the rates of charged-particle induced reactions, favoring  $(\gamma,n)$. Material accumulated at the $N=50$ neutron shell reaches stability fastest, while the temperature is still above 2~GK. Once stable isotopes are populated, $(p,\gamma)$ reactions below $N=50$ start to produce neutron-deficient isotopes, reaching as far as $^{84}$Zr. In the final abundances, however, we do not see the production of the lighter $p$-isotopes including $^{84}$Sr. This is because there is a continuous supply of free neutrons from $(\gamma,n)$ and $(p,n)$ reactions which allows for $(n,\gamma)$ reactions on the neutron deficient isotopes with $N<50$, which are quickly moved to the closed shell. As a consequence, $^{92}$Mo ($N=50$) is the lightest p-nucleus that is produced with this process. 

Due to the $r$-process operating during freeze-out from NSE, abundances in the iron-group are very low. Due to the high abundance of $^4$He, the $^{12}$C is still produced and proton-induced reactions make isotopes up to $^{38}$Ca, which cannot capture another proton and has a relatively long half-life of 444~ms. Since neutron-induced cross-sections scale with the mass number $A$, the heavier isotopes are much more likely to absorb the remaining free neutrons, such that $(n,p)$ reactions on the lighter isotopes are suppressed. As a result, the final abundances in Figure \ref{fig:n3_tracers} for track 146 show an accumulation of material around $A=38$ and very low mass fractions between $A=40$ and $A=90$. 

There are two peculiarities of the thermodynamic environment in our simulations that make this kind of process possible: First, the $r$~process operates in the presence of free neutrons and at high entropy \citep{Hoffman97}. \replaced{Secondly, some tracer particles do not follow monotonic expansion trajectories, i.e., after phases of very fast expansion that trigger the $r$~process, the temperature and density do not decrease exponentially, but instead, secondary phases of relatively moderate reheating and compression occur, allowing for two-step nucleosynthesis processes, as in the case of track 146. Such non-monotonic features of tracer particle trajectories can only be captured by multi-dimensional simulations. Similar effects have been seen, for example, in core-collapse supernova simulations \citep{Sieverding.Kresse.ea:2023}.} {Secondly, some tracer particles do not follow monotonic expansion trajectories in their final freeze out phase when nuclei are building up and are no longer dissociated.
For example, in track 146, the initially very fast expansion triggers the formation of heavy nuclei via the $r$~process but the temperature and density does not continue to decrease. Instead, secondary phases of reheating and compression occur, allowing for additional processing of the already formed $r$-process isotopes, thus constituting a two-step nucleosynthesis process.
Such non-monotonic features of tracer particle trajectories can only be captured by multi-dimensional simulations.
Modifications of well-known nucleosynthesis pathways by non-monotonic trajectories from multi-dimensional simulations have also been seen, for example, in core-collapse supernova simulations \citep{Sieverding.Kresse.ea:2023}.}

Of course not all trajectories produce heavy elements and escape from the gravitational pull of the neutron star. In fact we find that only about 25\% of tracers become unbound and make elements heavier than iron-group. This effective production efficiency is actually comparable to the most optimistic models
of neutron star mergers, which predict ejecta masses between $10^{-4}$ and 0.1  $M_\odot$ \citep{Cote18}. 
However, the uncertainty of extrapolating to long-term global behaviors within the CE environment,
along with the possibility of experiencing multiple or quasi-periodic accretion bursts, 
makes absolute predictions of the total mass conversion difficult. In the context of a brief single
burst event exemplified by the $\dot m = 10^{10}$ model, and because we have distributed
tracer particles to uniformly sample the (local) spherical volume, we can approximate the total mass of
accreted material converted to heavy elements and (potentially) injected into the IGM as
$\delta M = \dot{M} ~x_f ~\tau_{esc} \approx 10^{-7} M_\odot$, where $\dot{M} = 30 ~M_\odot yr^{-1}$, $x_f$ is the fraction of tracers making elements with $A > 60$, and $\tau_{esc}$ is the shock relaxation time.

\begin{figure}
\hspace{0.5in}\includegraphics[width=0.8\textwidth]{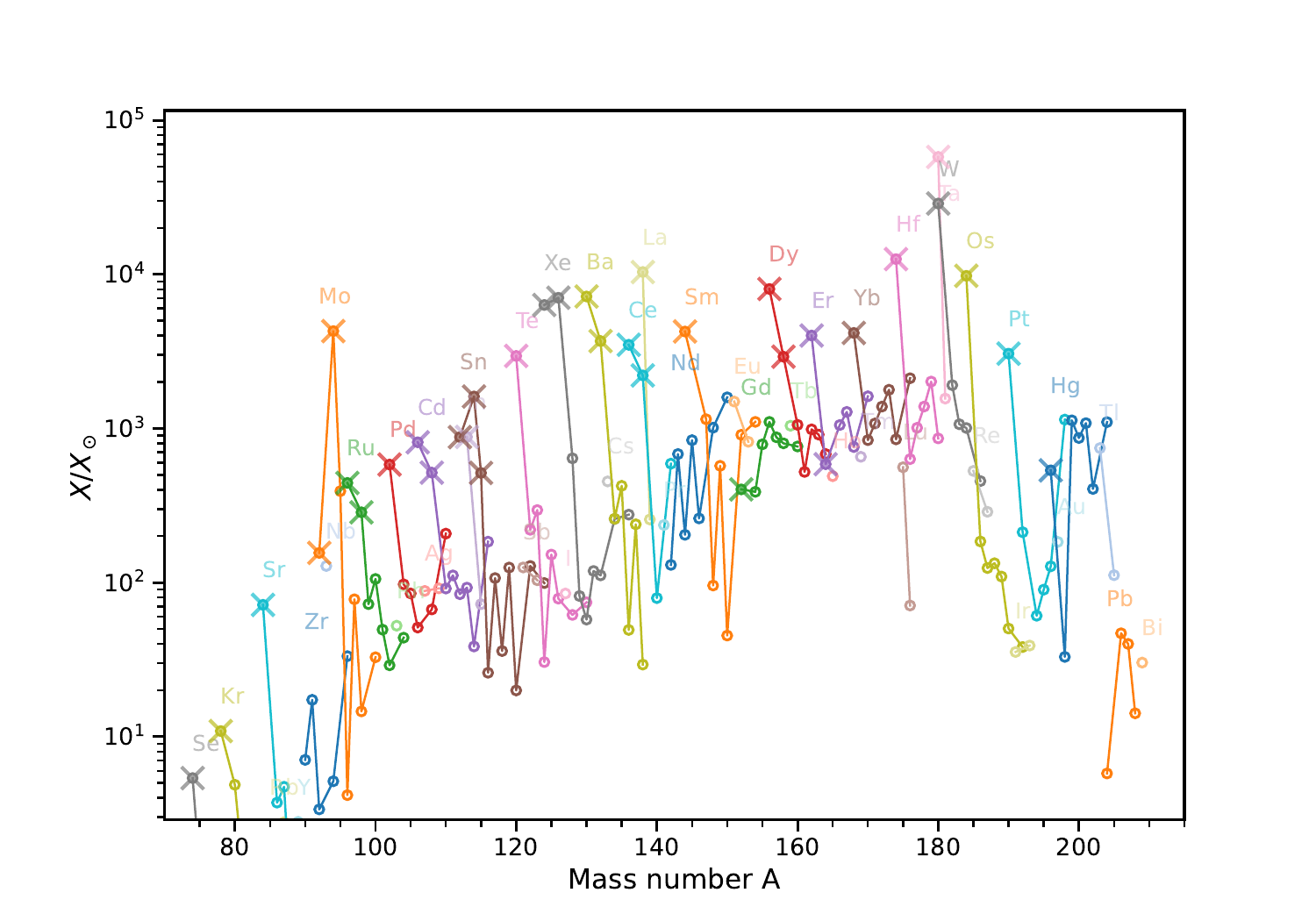}
\caption{
Production factors based on the cumulative mass fractions from Figure \ref{fig:solar_alltrks} summed over all escaping tracer particles from the $\dot m = 10^{10}$ model. The $p$-nuclei, i.e., isotopes that cannot be made by neutron captures are marked with crosses. Even though just a few tracer particles contribute to the $p$-isotopes, the very small solar abundance leads to large overproduction factors.
}
\label{fig:solar_alltrks_ovr}
\end{figure}

One final point we wish to make concerns over-production factors of the CE aggregate, i.e., the total mass fractions relative to the corresponding fractions in the solar system. We plot in Figure \ref{fig:solar_alltrks} the mass fractions summed over unbound tracers from the $\dot m = 10^{10}$ model, accounting for all nuclear products contributed by every tracer on an escape trajectory. We also superimpose solar system abundances from \cite{Asplund.Grevesse.ea:2009} and \cite{Goriely:1999}, normalized to match the iron abundance from the numerical model. 
\replaced{Notice the rough similarity in the pattern of atomic elements with $A > 100$, and the overabundance or production factors ranging between $10$ to $10^5$ times higher than solar.}{As noted by \citet{Meyer02}, the abundance pattern of a high-entropy $r$ process differs significantly from the solar pattern.}
In Figure \ref{fig:solar_alltrks_ovr} we show the actual overproduction factors relative to the data from \cite{Asplund.Grevesse.ea:2009}. In spite of the small number of tracer particles that allow the production of $p$-nuclei via the mechanism described above for track 146, the impact of this mechanism can be large on average because of the extremely small solar abundances of $p$-isotopes. With this simple average over tracer particles, the $p$-nuclei turn out to dominate the final composition. It is interesting that the production factors of the $p$-nuclei are high and relatively flat across the whole range from Xe to Os, and includes Mo, which makes CE environments possible sites to produce most of the $p$-isotopes found in the solar system.

\section{Summary}
\label{sec:summary}

Although the kilonova associated with GW170817 has given strong observational evidence for one site of the $r$-process, many questions regarding the origin of heavy elements in the universe remain uncertain, including the large scatter of $r$-process abundances in metal-poor stars as well as the $p$-nuclei in the solar system. Here we have explored the possibility of observing $r$-process nucleosynthesis and $p$ nuclei in a Common Envelope binary consisting of a neutron star accretor and a red giant providing helium-rich fuel. At the NS surface, gas accreting from the red giant companion drives an outgoing convectively unstable shock creating an environment where nucleons evolve through repeated cycles of photo-disintegration by convective heating, and multi-step $r$, $p$ and $\gamma$ process nucleosynthesis by convective cooling. Material is eventually ejected in a final freezeout with a distribution of heavy elements that depends strongly on the accretion rate, among other factors. 

Our models resolve the nuclear environment close to (within a thousand kilometers of) the neutron star, including the thin neutrino-sphere atop the neutron surface and the convective instability developing behind the accretion shock. These conditions generate significantly lower luminosities or neutrino fluxes compared to supernova events that are more commonly associated with the $r$-process. However we find that at sufficiently high (hyper-Eddington) accretion rates, our models develop extremely high entropies. This, together with the short (millisecond) expansion timescales experienced by fluid parcels on their final escape trajectory, promotes the right conditions for the $r$-process through a unique, neutron depleted, disequilibrium mechanism. 

It is generally assumed that neutron-rich environments are required for the $r$-process. Tracer histories extracted from our models instead are slightly proton-rich, beginning with $Y_e \sim 0.505$ as fluid enters the computational domain and either remains that way until the tracers are ejected, or increases slightly. Heavy element nucleosynthesis is possible in this situation because the high entropy environment of the CE prevents all the neutrons from being absorbed into alpha particles during the rapid expansion phase, leaving a significant neutron-to-seed ratio to facilitate the $r$-process. This is essentially a disequilibrium effect easily missed had we applied NSE too prematurely in our models (at temperatures below 8 GK).

We find that trajectories producing the heaviest elements do so through the $r$-process on the neutron-rich side of stability, and those that wound up with lower mean atomic masses made proton-rich isotopes through a combination of NSE freezeout and multi-step reheating effects that drive neutron-rich elements towards stability via proton-induced reactions. Taken together, the abundances of all ejected elements yield a distribution that is dominated by $p$-nuclei with high overproduction factors. More than 25\% of tracers sampling the accretion domain escape the binding energy of the NS after producing significant amounts of heavy elements ($A>100$) over a dynamical relaxation timescale (when the shock cools to temperatures incapable of sustaining further nuclear yields). The remainder of material either falls to the neutron star surface as disassembled neutrons and protons, or remains bound to the star.

Although we have neglected effects of magnetic fields attached to the neutron star (justified by the overwhelming  accretion rates), it is possible that fields embedded with the accreting plasma may affect the development of turbulence and convective turnover. We leave the inclusion of magnetic fields for future work, along with considerations of accreting material with different initial atomic compositions and arbitrary (non-spherical) flow configurations.

\begin{acknowledgments}

This work was performed in part under the auspices of the U.S. Department of Energy by 
Lawrence Livermore National Laboratory under Contract DE-AC52-07NA27344. 

\end{acknowledgments}

\software{ Cosmos++ \citep{Anninos05,Fragile14,Anninos17,Anninos20,Roth22}, PRISM \citep{Sprouse.Mumpower.ea:2021,Sprouse2020}, Winnet \citep{Reichert23}, XNet \citep{Hix.Meyer:2006} }

\bibliography{refs_hyper,refs_nucleo}{}
\bibliographystyle{aasjournal}


\end{document}